\documentclass[runningheads]{llncs}

\usepackage[utf8]{inputenc}
\usepackage{CJKutf8}
\usepackage{natbib}
\usepackage{graphicx}
\usepackage[hyphens]{url}  
\usepackage{hyperref}
\usepackage{pdflscape}
\usepackage{longtable}
\usepackage{array}
\usepackage[a4paper, left=1.cm, right=1.8cm, top=2.5cm, bottom=2.5cm]{geometry}
\usepackage{booktabs}
\usepackage{multirow}
\usepackage{adjustbox}
\usepackage{enumitem}

\usepackage{tabularx}
\usepackage{caption}
\usepackage{chngcntr}
\counterwithin{table}{section}
\usepackage[toc,page]{appendix}

\hypersetup{breaklinks=true}

\makeatletter
\renewcommand\subsubsection{\@startsection{subsubsection}{3}{\z@}%
  {-18pt \@plus -4pt \@minus -4pt}%
  {0.5em \@plus 0.22em \@minus 0.1em}%
  {\normalfont\normalsize\bfseries}}
\makeatother

\begin{document}
\title{Discovering Self-Regulated Learning Patterns in Chatbot-Powered Education Environment}
%
%
\author{
Yilin Lyu\inst{1}\orcidID{0009-0009-3808-3802} \and
Ren Ding\inst{2}\orcidID{0009-0004-1784-6894}
}

\authorrunning{Y. Lyu et al.}

\institute{
School of AI and Liberal Arts, BNBU-HKBU United International College, Zhuhai, China\\
\email{lyuyilin@uic.edu.cn} \and
School of Computing and Information Systems, The University of Melbourne, Grattan Street, Parkville, Victoria 3010, Australia\\
\email{dr2767192219@gmail.com}
}

\maketitle              
\begin{abstract}
The increasing adoption of generative AI (GenAI) tools such as chatbot in education presents new opportunities to support students’ self-regulated learning (SRL), but also raises concerns about how learners actually engage in planning, executing, and reflection when learning with a chatbot. While SRL is typically conceptualized as a sequential process, little is known about how it unfolds during real-world student–chatbot interactions. To explore this, we proposed Gen-SRL, an annotation schema to categorize student prompts into 16 microlevel actions across 4 macrolevel phases. Using the proposed schema, we annotated 212 chatbot interactions from a real-world English writing task. We then performed frequency analysis and process mining (PM) techniques to discover SRL patterns in depth. Our results revealed that students’ SRL behaviours were imbalanced, with over 82\% of actions focused on task execution and limited engagement in planning and reflection. In addition, the process analysis showed nonsequential regulation patterns. Our findings suggest that classical SRL theories cannot fully capture the dynamic SRL patterns that emerge during chatbot interactions. Furthermore, we highlight the importance of designing adaptive and personalized scaffolds that respond to students' dynamic behaviours in chatbot-powered contexts. More importantly, this study offers a new perspective for advancing SRL research and suggest directions for developing chatbots that better support self-regulation.

\end{abstract}

\section{Introduction}
\label{sec:introduction}
Learning analytics (LA) is the systematic collection and analysis of learner-related data to better understand and improve both learning processes and the environments in which they take place~\cite{long2014penetrating}. As an interdisciplinary research area, LA aims to use data to gain deeper insights into student behaviour and improve educational interventions. Within the field of LA, self-regulated learning (SRL) is often used as a theoretical lens to interpret and categorise learner behaviours. SRL describes how learners actively set goals, monitor progress, adapt their strategies, and reflect on outcomes in order to regulate their learning~\cite{pintrich2000role,zimmerman2000attaining}. It emphasizes the active role of learners in managing their learning path and is widely regarded as a key determinant of academic success~\cite{panadero2017}. Its importance has been further highlighted in the post-COVID-19 era, where the shift to online and hybrid learning environments has required students to engage in self-regulation without continuous teacher supervision~\cite{biwer2021,sulisworo2020students}. As SRL research has progressed, a variety of theoretical models have been proposed~\cite{zimmerman2000attaining,pintrich2000role}. These models typically conceptualize SRL as a sequential process through the orderly execution of three key phases: (1) Planning, (2) Performance, and (3) Reflection. To apply theory into practice, SRL has been measured using self-report questionnaires such as the MSLQ~\cite{pintrich1991manual}. However, such tools capture only a static view of SRL and students always misreport it due to recall bias~\cite{zimmerman2000attaining,rovers2019granularity}. Driven by the growth of online learning environments, researchers have increasingly turned to using digital footprint, such as clickstreams or page viewing histories from the Learning Management System (LMS), to analyse SRL behaviours~\cite{armas2024dusting,tempelaar2024understanding}. Among these digital-based measurement methods, microlevel process analysis has emerged as a key approach that maps learner's learning behaviours to meaningful SRL actions~\cite{greene2009macro} (as detailed in Section~\ref{lr:challenges in measuring srl}). However, this method still faces limitations when applied to digital data, which often contain ambiguous semantic data or noisy data~\cite{bannert2014process,armas2024dusting}.

In recent years, Generative AI (GenAI) has been increasingly integrated into educational contexts, particularly through the use of GenAI-powered chatbot~\cite{clarizia2018chatbot,lyu2024discussion}. This chatbot supports interactive conversations by answering questions, explaining concepts, and offering real-time guidance~\cite{kuhail2023interacting}. In contrast to previous learning environments, chatbot dialogues serve as “open traces of thought”, providing greater semantic transparency and thus making self-regulation processes easier to capture~\cite{xu2025enhancing}. At the same time, Bushuyev et al. describe today's GenAI-mediated learning landscape as a brittle, anxious, nonlinear, and incomprehensible (BANI) world~\cite{bushuyev2023bani}, highlighting GenAI is reshaping education as students increasingly depend on these tools. In such environments, learners may regulate their behaviour in more fluid and context-dependent ways~\cite{chang2023educational}.
However, limited research has examined how SRL behaviours emerge in chatbot-powered settings. To address this gap, this study proposes the following two research questions.

\vspace{1em}
\noindent\textbf{RQ1.}\hspace{0.5em}
\textit{How can we process and interpret student–chatbot interactions from the perspective of SRL? }
This question aims to fill a current gap in the measurement of SRL behaviours in chatbot-based learning environments. Although GenAI-powered chatbots are widely used, there are no established methods for identifying and interpreting SRL behaviours from chatbot conversations. This question focuses on developing an annotation schema to translate student-chatbot interactions into meaningful SRL actions for future analysis. It also underpins RQ2, as the discovery of SRL patterns depends first on addressing this question.

\noindent\textbf{RQ2.}\hspace{0.5em}
\textit{What are students' actual SRL behaviour patterns in chatbot-powered learning environments?} Based on the results of RQ1, this question aims to explore SRL behaviour patterns in chatbot-mediated contexts, examine their deviations from traditional models. The findings may guide targeted interventions to enhance self-regulation in AI-supported learning.

\vspace{1em}

This study addresses these questions through three main contributions.

\begin{enumerate}[label=\arabic*)]
\item Schema Formulation Dimension: This study proposes Gen-SRL, a novel annotation schema for measuring SRL behaviours in chatbot-mediated learning environments (to respond to RQ1).
\item Discovery Dimension: This study is the first to apply process mining (PM) as the primary analytical method to model and visualize learners' SRL paths using real-world student–chatbot interaction data (to respond to RQ1).
\item Analytic Dimension: This study identifies imbalanced SRL behaviour distributions and nonsequential patterns in chatbot-based learning, challenging classical SRL assumptions and providing insights for adaptive intervention design in GenAI-powered education (to respond to RQ2).
\end{enumerate}

The remainder of this paper is organized as follows. Section~\ref{sec:related work} reviews existing SRL theories, measurement methods, expected SRL patterns, and the use of PM in SRL research. Section~\ref{sec:methodology} details the formulation of Gen-SRL annotation schema and its implementation to real-world dataset. Section~\ref{sec:results} presents the results and Section~\ref{sec:discussion} provides an in-depth analysis of the observed SRL patterns and evaluates their implications. Section~\ref{sec:limitations and future work} outlines key limitations and proposes directions for future research. Finally, Section~\ref{sec:conclusion} concludes the paper.

\section{Literature Review}
\label{sec:related work}
\subsection{Self-Regulated Learning}
SRL refers to how learners actively control their learning process to achieve learning goals. It emphasizes the ability of learners to direct and manage their cognitive, motivational, and behavioural processes to achieve learning goals~\cite{panadero2017review}. SRL yields significant psychological and socio-cognitive benefits, positively influencing academic achievement, motivation, and self-perception~\cite{zimmerman1989social,she2023learning}. Through strategies such as goal setting, progress monitoring, and feedback integration, SRL fosters metacognitive awareness, improves self-efficacy, and strengthens emotional resilience~\cite{luo2024effectiveness}. These capabilities are fundamental to lifelong learning, personal growth, and adaptability in diverse contexts. The importance of SRL has become even more apparent in digitally mediated learning scenarios, especially during the COVID-19 pandemic, where students and educators had to rely on distance learning technologies such as online learning systems~\cite{biwer2021,sulisworo2020students}. These transitions highlighted the urgent need for learners to self-regulate without continuous teacher supervision and underscored SRL as a core competency in modern education.

Over the past decades, extensive research has explored various theoretical models of SRL, aiming to unpack the psychological mechanisms behind effective self-regulation~\cite{pintrich2000role,puustinen2001models,zimmerman1989social,zimmerman2000attaining}. Zimmerman's cyclical model~\cite{zimmerman2000attaining} outlines three key phases: forethought, performance, and self-reflection. Complementing this, Pintrich’s four-phase model~\cite{pintrich2000role} subdivides the performance phase into monitoring and control, emphasizing their interdependence through a feedback loop where learners monitor their progress, adjust strategies accordingly and then re-monitor the results of these adjustments~\cite{SchunkZimmerman2012}. This structure offers a more nuanced understanding of self-regulation behaviour. Other notable frameworks include Boekaerts and Cascallar’s Dual Processing Model~\cite{boekaerts2006far}, Winne and Hadwin's COPES Model~\cite{winne2011cognitive}, and the MASRL Model by Efklides~\cite{efklides2011interactions}. In addition to outlining high-level phase structures, these SRL models also detail specific subprocesses or actions within each phase. For example, Pintrich's model identifies subprocesses such as goal setting, self-observation, and task evaluation~\cite{pintrich2000role}. These theoretical foundations have laid the foundation for measuring and identifying SRL behaviours in authentic learning environments.

\subsection{SRL Patterns in Classical Theories}
\label{subsec:srl_patterns}    
Classical models of SRL consistently describe SRL as a balanced and sequential process~\cite{saint2020combining}. Firstly, learners are expected to engage equitably in all phases, rather than focus disproportionately on any single phase~\cite{pintrich2000role,boekaerts2000self}. Zimmerman emphasized that effective self-regulators evenly traverse the forethought, performance, and self-reflection phases throughout learning, rather than favoring one over others~\cite{zimmerman2000attaining}. Similarly, Pintrich argued that failing to balance any single phase would compromise the integrity of the overall regulation process~\cite{pintrich2000role}.

\begin{figure}[htbp]
\centering
\includegraphics[width=0.7\linewidth]{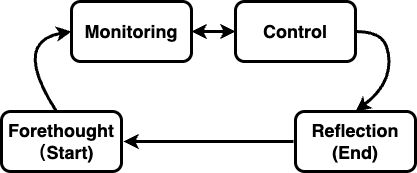}
\caption{A reproduction of the theoretical SRL model, based on Pintrich's four-phase model~\cite{pintrich2004}.}
\label{fig:traditional_srl_model}
\end{figure}

In addition to balanced engagement, the definition of SRL models as a sequential learning cycle that unfolds over time is well-established~\cite{butler1995feedback,molenaar2014advances}. In this view, learners are expected to progress in a fixed order from planning to execution, reflection, and back to planning~\cite{pintrich2000role}. We reproduced this process based on Pintrich's four-phase model as shown in Figure~\ref{fig:traditional_srl_model}, which illustrates the idealized sequence:
\textit{Forethought $\rightarrow$ Monitoring $\leftrightarrow$ Control $\rightarrow$ Reflection $\rightarrow$ Forethought}.
The bidirectional arrow between \textit{Monitoring} and \textit{Control} highlights the iterative nature of in-task adjustments, which reinforce rather than disrupt the overall sequential cycle~\cite{hacker1998metacognition}.
. Zimmerman also emphasized that self-regulated learning pattern unfolds through an ordered sequence of regulatory activities~\cite{zimmerman1989social,zimmerman2000attaining}. Clearly et al. explicitly uses the term "sequential phases of regulation" to describe the sequential and cyclical nature of regulated learning~\cite{cleary2012assessing}.

To address \textbf{RQ2}, we draw on classical SRL patterns~\cite{pintrich2000role,zimmerman2000attaining} and hypothesize that students' SRL behaviours in chatbot-powered contexts will exhibit balanced and sequential engagement, as stated in the two hypotheses below.

\textbf{H1.} \textit{Students in chatbot-powered learning environments are expected to exhibit a balanced distribution of SRL behaviours.}

\textbf{H2.} \textit{In chatbot-powered learning environments, students' SRL patterns are assumed to follow a sequential structure.}

These hypotheses reflect theoretical expectations from established SRL models. In this study, we empirically test whether these assumptions hold in GenAI-supported contexts. 

\subsection{SRL Measurement}
\label{lr:challenges in measuring srl}
Measuring SRL has long posed a challenge for researchers~\cite{molenaar2023measuring}. Traditional approaches rely primarily on self-report instruments, such as the Motivated Strategies for Learning Questionnaire (MSLQ)\cite{pintrich1991manual}, which asks learners to reflect on their learning strategies and motivation after tasks. While informative, such methods offer only static snapshots of SRL and are prone to recall bias and social desirability effects, making them inadequate to capture the dynamic nature of SRL\cite{zimmerman2000attaining,rovers2019granularity}.

With the widespread use of LMSs and online learning platforms, researchers have increasingly turned to behavioural data captured in digital learning environments~\cite{araka2020research}. Rather than relying on the learners' retrospective reports, digital trace data, such as clickstreams, resource access logs or chat histories, can objectively reveal how SRL unfolds in real time~\cite{winters2008self,tempelaar2024understanding}. Among the various trace-based analytical approaches, microlevel process analysis has emerged as one of the most effective methods for mapping raw trace data to meaningful SRL constructs~\cite{greene2009macro}. 

In this approach, a coding scheme (i.e., the annotation schema in this study) serves as the key mechanism to map observable trace behaviors to microlevel SRL actions, which are further organized into broader macrolevel phases such as planning, performance and reflection~\cite{zimmerman2000attaining}. For instance, dragging a resource into a goal list may be coded as \textit{Goal Setting} within the planning phase. Table~\ref{tab:srl_micro} presents a hierarchical view of such mappings. Early work by Siadaty et al.\cite{siadaty2016trace} applied this structure to LMS trace data, defining fine-grained actions through manual coding of clicks, scrolls, and ratings\cite{cleary2001,osakwe2024measurement}.

However, this method is highly sensitive to context~\cite{winne2017learning}. The diversity of LMS platforms often necessitates custom-built mapping schemes that are time-consuming and subjective. Moreover, LMS interaction logs can contain noise (e.g., accidental clicks) and lack semantic clarity~\cite{rienties2016implementing}. For instance, an accidental click on a quick glance at a reading page may be logged as a valid action, even without real engagement. These constraints have motivated a shift toward richer and more meaningful behavioural traces. The rapid rise of GenAI tools, such as chatbots, is transforming educational practices and offering new opportunities for SRL measurement. Unlike conventional LMS logs, this setting offers a more transparent behavioural context that allows a more effective capture and analysis of learners' SRL behaviours~\cite{azevedo2019analyzing}. While prior studies have explored students' intentions in student-chatbot interactions~\cite{2025zixin,han-etal-2024-recipe4u-student}, none have systematically processed and analysed these dialogues from the perspective of SRL. To address this gap and answer RQ1, we introduce an annotation schema to process chatbot dialogues (see Section~\ref{sec:methodology}).

\begin{table}[htbp]
\centering
\caption{Hierarchical SRL framework with macrolevel phases, microlevel actions, and specific learner behaviours (e.g., learners' operations in LMS), adapted from~\cite{siadaty2016}.}
\label{tab:srl_micro}
\begin{tabular}{p{4.1cm} p{4.2cm} p{7cm}}
\toprule
\textbf{Macrolevel Phases} & \textbf{Microlevel Actions} & \textbf{Specific Learner Actions} \\
\midrule
\multirow{3}{*}{Planning} 
  & Task Analysis & Clicking on different competences under duties or projects \\
  & Goal Setting & Dragging and dropping an available competence to a new/existing learning goal \\
  & Making Personal Plans & Choosing an available learning path as the path \\
\midrule
\multirow{2}{*}{Performance} 
  & Working on the Task & Marking a learning goal as “completed” \\
  & Strategy Adjustment & Updating the properties of a learning activity \\
\midrule
\multirow{2}{*}{Reflection} 
  & Self-Evaluation & Rating a learning path, learning activity, or knowledge asset \\
  & Revision & Adding a comment for a competence, learning path, or learning activity \\
\bottomrule
\end{tabular}
\end{table}

\subsection{Process Mining in SRL}
Process mining, initially developed within the field of Business Process Management (BPM), aims to automatically discover, visualize, and analyse hidden process structures from event logs \cite{van2016data,van2004workflow}. Given its strength in handling sequential event data, PM has expanded into educational contexts, forming the subfield known as Educational Process Mining (EPM). EPM is increasingly used to analyse student learning behaviours captured on digital platforms such as LMS~\cite{abel2023uncovering,lyu2024mamba}.

Bannert et al. \cite{bannert2014process} highlighted PM as a promising methodological approach to investigate SRL. PM methods enable the analysis of sequential relationships by incorporating insights from frequency, temporal order, and probabilistic patterns. Specifically, PM techniques transform students' activity data from computer-based learning environments into visualized process models \cite{romero2010educational}, effectively capturing the workflow structures inherent in learning activities \cite{van2004workflow}. Moreover, different PM algorithms explicitly address noise in the education data (i.e., infrequent or exceptional behaviours). Thus, PM is particularly suitable for examining SRL behavioural patterns informed by process-oriented theoretical assumptions \cite{sonnenberg2015discovering}.


Extensive research has applied PM to reveal differences in SRL patterns among learners. Bannert et al. \cite{bannert2014process} employed the Fuzzy Miner algorithm on think-aloud data, revealing that high-achieving students often planned before learning, engaged in deeper cognitive strategies such as elaboration and monitoring during learning, and conducted reflective evaluations post-task. In contrast, low-achieving students predominantly used superficial cognitive strategies, such as repetition, and rarely participated in reflective activities. Similarly, Leung \cite{leung2022mapping} and Cerezo et al. \cite{cerezo2020process} applied PM techniques to LMS log data, demonstrating that high-performing students consistently exhibited coherent monitoring and goal-setting strategies. Saint et al. \cite{saint2020tracesrl} utilized microlevel process analysis and stochastic PM algorithms in an online learning environment, confirming that more successful learners consistently demonstrated a higher frequency of SRL behaviours than their less successful counterparts. Current studies remain limited to traditional or online settings and have not applied PM to GenAI-mediated learning environments~\cite{molenaar2023measuring}. This study is among the first to systematically explore SRL behaviours in chatbot-powered contexts using PM techniques.

\section{Methodology}
\label{sec:methodology}

\begin{figure}[htbp]
\centering
\includegraphics[width=0.86\linewidth]{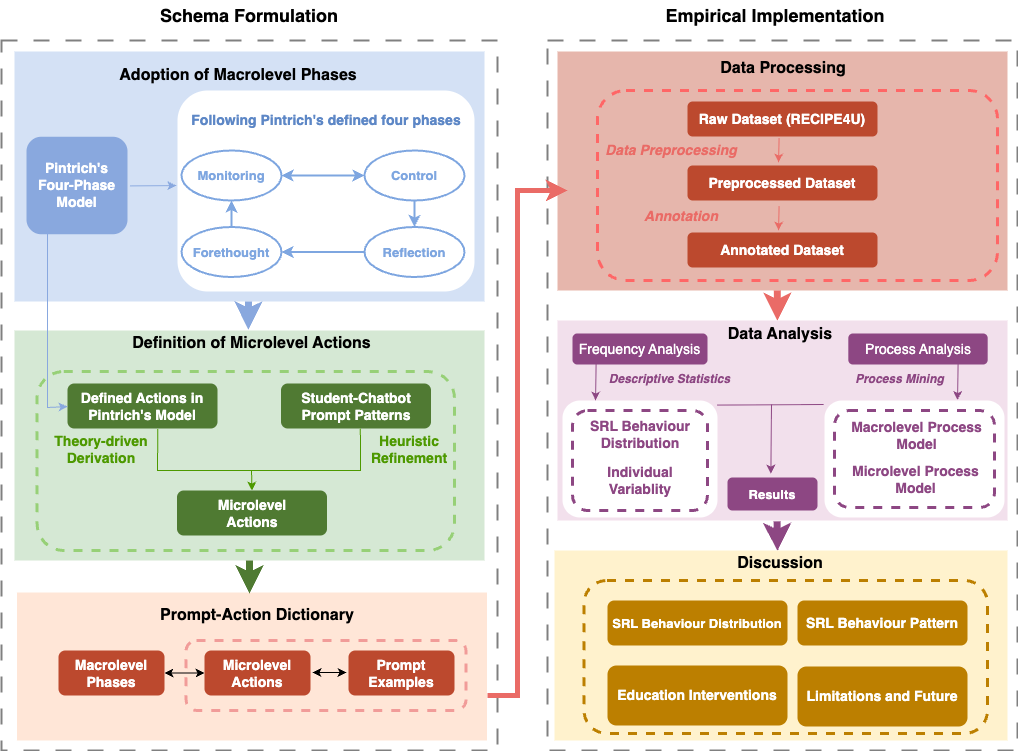}
\caption{Overview of the proposed methodology, consisting of schema formulation and empirical implementation stages.}
\label{fig:gen-srl-method}
\end{figure}

To address \textbf{RQ1}, we designed a two-part methodology consisting of (1) the formulation of a Gen-SRL annotation schema and (2) its empirical implementation. The first part details the development of our annotation schema grounded in existing theory and tailored to GenAI-based interactions. The second part demonstrates a proof-of-concept, including dataset preparation, annotation procedures, and data analysis methods. Figure~\ref{fig:gen-srl-method} provides an overview of our proposed methodology. 

\subsection{Schema Formulation}
\label{subsec: Gen-SRL Framework}
To systematically capture and analyse students' SRL behaviours in GenAI-mediated contexts, we developed Gen-SRL, a hierarchical annotation schema grounded in microlevel process analysis (as introduced in Section~\ref{lr:challenges in measuring srl}). It comprises three core components:
(1) 4 macrolevel SRL phases,
(2) 16 microlevel SRL actions, and
(3) a prompt–action dictionary that maps student utterances to specific SRL behaviours.
Table~\ref{tab:srl-framework} presents the structure of the Gen-SRL.

We first examined existing SRL theories and adopted Pintrich's four-phase model~\cite{pintrich2000role}, which comprises forethought, monitoring, control, and reflection. Compared to classical three-phase models that merge monitoring and control into a single phase, Pintrich's model provides a clearer and more granular structure. This makes it suitable for identifying and annotating dynamic learning behaviours in GenAI-supported contexts~\cite{zimmerman2000attaining,mooij2014self,Puustinen2001,winters2008self}. Within each phase, we identified 16 microlevel SRL actions through a combination of theory-driven derivation and heuristic refinement. Theory-driven actions, such as \textit{Define Problem} and \textit{Self-Evaluate}, were directly derived from the predefined actions in Pintrich's model~\cite{pintrich2000role}. In contrast, heuristic refinement involved adapting or extending existing theoretical ideas to better fit GenAI-mediated contexts. For example, the predefined action \textit{Help seeking behaviour} in Pintrich's model was subdivided into \textit{Request Help-Check}, \textit{Request Help-Instrumental}, and \textit{Request Help-Executive} based on observed prompt patterns that reflect more concrete learner intentions. Similarly, \textit{Refine Prompt} builds on the theoretical action of \textit{"use of cognitive strategies"}, but is heuristically refined to capture a behaviour that is specific to chatbot-based learning. This approach ensures that our schema remains grounded in established theory while accommodating the nuances of GenAI-mediated learning environments. The theoretical or heuristic origin of all actions is marked in Table~\ref{tab:srl-framework}. In addition, each action was further classified as cognitive or metacognitive based on its regulatory function \cite{flavell1979metacognition}. Note that this classification depends primarily on whether the action directly supports task execution (cognitive) or involves conscious regulation and monitoring of learning (metacognitive). All actions are summarized in Table~\ref{tab:srl-framework}, then we describe them by phase below.

\textit{Forethought} phase involves all preparatory processes before learners take action. It includes analyzing the learning problem and setting relevant learning goals. Problem analysis and setting goals are considered the starting points of SRL \cite{maghfiroh2017problem,Zimmerman2008}. Note that \textit{Forethought} is typically regarded as a metacognitive phase, and its actions are generally classified as metacognitive behaviours~\cite{SchunkZimmerman2012}. Based on the actions defined in Pintrich's model, we adopted two metacognitive actions in this phase: \textit{Define Problem} and \textit{Set Goals}, as detailed in Table~\ref{tab:srl-framework}.

\textit{Monitoring} phase involves continuously monitoring the progress during the task. Together with the control phase, this forms a feedback loop \cite{SchunkZimmerman2012}, in which learners monitor their understanding and adjust their learning actions accordingly. We define two monitoring actions, i.e., \textit{Check Understanding-Basic} and \textit{Check Understanding-Deep}, which capture learners' evaluation of their understanding of AI responses. The distinction lies in the level of thinking, with \textit{Basic} reflecting straightforward understanding and \textit{Deep} involving more analytical and reflective thought. Both are classified as metacognitive behaviours, as they reflect self-monitoring and awareness of learning processes \cite{flavell1979metacognition}.

\textit{Control} phase involves adjusting learning behaviours based on self-monitoring. We identified eight actions in this phase (see Table~\ref{tab:srl-framework}), including help-seeking (\textit{Request Help Instrumental}, \textit{Request Help Check}, \textit{Request Help Executive}), clarification (\textit{Seek Clarificatio-Basic}, \textit{Seek Clarification-Deep}), information retrieval (\textit{Access Information}), and strategy refinement (\textit{Correct Answer}, \textit{Refine Prompt})\cite{ekin2023prompt}. Among these, three actions are metacognitive: \textit{Refine Prompt} reflects learners' effort to improve their input after recognizing its inadequacy, \textit{Correct Answer} involves verifying and correcting AI responses, and \textit{Seek Clarification Deep} aims to deepen understanding through strategic questioning. In contrast, actions such as \textit{Access Information} and \textit{Seek Clarification Basic} directly support task completion and are therefore cognitive. Although metacognitive behaviours exist, the control phase is overall dominated by cognitive strategies, consistent with prior SRL theories \cite{zimmerman2000attaining,pintrich2000role}.

\textit{Reflection} phase, similar to \textit{Forethought}, is widely considered the metacognitive phase. It involves evaluating one's performance and considering future directions after task completion. This phase focuses on reviewing previous strategies and outcomes to generate metacognitive insights that inform future self-regulated learning \cite{winters2008self}. Based on the definitions in Pintrich's model, we defined three actions in this phase: \textit{Self-Evaluate}, \textit{Summary Learning}, and \textit{Plan Next Step}, all of which are classified as metacognitive.

In addition, to improve the measurement precision, we introduced a non-SRL phase that includes the action named \textit{Acknowledgment} to capture prompts such as "Thank you" or "Good morning". These prompts do not contribute to the regulation process and are excluded from the subsequent analysis \cite{roll2011helpseeking,winne2010framework}.

To apply this schema to real-world chatbot conversations, we followed the microlevel process analysis method outlined in Section~\ref{lr:challenges in measuring srl} and constructed a prompt-action dictionary that maps representative student prompts to predefined SRL actions through heuristic matching. This dictionary (often termed a coding scheme) serves as an operational bridge between abstract theoretical constructs and observable student–chatbot interaction data, forming the core of the microlevel process analysis method discussed in Section~\ref{lr:challenges in measuring srl}. Selected prompts examples are provided in Table~\ref{tab:srl-framework}, and the complete dictionary is available in Appendix A. This hierarchical annotation schema supports subsequent macrolevel and microlevel SRL analysis in actual chatbot-powered settings. 

\newgeometry{left=1.8cm, right=1.8cm, top=0.5cm, bottom=0.5cm}
\begin{table}[b]
\centering
\begin{adjustbox}{angle=90, left, width=\textheight, height=\textwidth, keepaspectratio}
\renewcommand{\arraystretch}{1.2}
\fontsize{8}{8}\selectfont
\begin{minipage}{\linewidth} %
\centering
\caption{The Gen-SRL annotation schema used for annotating SRL behaviors in chatbot-mediated learning, including macrolevel phases, microlevel actions, codes, types (cognitive/metacognitive), origin (theoretical/heuristic), descriptions, and representative prompts.}
\label{tab:srl-framework}
\begin{tabular}{%
  >{\raggedright\arraybackslash}p{1.7cm}
  >{\raggedright\arraybackslash}p{2.2cm}
  >{\centering\arraybackslash}p{1.2cm}
  >{\centering\arraybackslash}p{2.0cm}
  >{\centering\arraybackslash}p{1.6cm}
  >{\raggedright\arraybackslash}p{6.5cm}
  >{\raggedright\arraybackslash}p{5.5cm}
}
\toprule
Macrolevel Phase&Microlevel Action&Code&Type&Origin&Description&Example \\ \midrule
Forethought  &Define Problem&F.DP&Metacognitive&Theoretical&The learner identifies problems or learning needs before task.&1. Today I need to solve ...\newline
2. I have a problem about ... \\
&Set Goals&F.SG&Metacognitive&Theoretical&The learner determines a specific outcome they wish to achieve.&1. I want to understand/write ... \\ \midrule
Monitoring   &Check Understanding - Basic&M.CU(B)&Metacognitive&Heuristic&The learner verifies their basic understanding by paraphrasing, summarizing or questioning AI's response.&1. I see. That means ...;\newline 
2. So are you saying ...?\\
&Check Understanding - Deep&M.CU(D)&Metacognitive&Heuristic&The learner uses higher-order skills to verify or challenge their understanding, such as through example testing, counter-example testing, analogies, concept comparison, or transfer learning.&1. If … is …, then can it still be considered …?\newline
2. So this is like …, right?\newline
3. Would this also apply if I were doing … instead of …?\newline
4. So it is different from …? \\ \midrule
Control  &Seek Clarification - Basic&C.SC(B)&Cognitive&Heuristic&The learner asks straightforward questions to clarify basic concepts.&1. Can you clarify the concept you just mentioned?\newline
2. Can you make it easier to understand?
 \\
&Seek Clarification - Deep&C.SC(D)&Metacognitive&Heuristic&The learner asks in-depth or follow-up questions to thoroughly explore a concept, such as probing causes, comparing alternatives, or requesting examples.&1. Can you give an example for this concept?\newline
2. Why is this option better than the others?\newline
3. What is the difference between concept A and B? \\
&Request Help - Instrumental&C.RH(I)&Cognitive&Heuristic&The learner uses chatbot as a task coach to seek methods or strategies for completing the task, such as consulting writing structure, logic or organization.&1. How can I make this paragraph longer?;\newline
2. How can I re-structure the following sentence?;\newline
3. Can you suggest any improvements on ...? \\
&Request Help - Executive&C.RH(E)&Cognitive&Heuristic&The learner asks the AI to directly complete a task for them.&1. Please generate a new introduction for this essay.;\newline
2. Write me a summary paragraph. \\
&Request Help - Check&C.RH(C)&Cognitive&Heuristic&The learner requests the AI to review, check, or revise a task they have already done&1. Can you help me revise/check ...?;\newline
2. How would you rate ...? \\
&Access Information&C.AI&Cognitive&Heuristic&The learner treats chatbot as a tool to retrieve factual information, translate, and count words or general ChatGPT-related questions without further processing.
&1. What is ...(non-concept question)?\newline
2. Can you give some references about ...?;\newline
3. Can you translate it into English?;\newline
4. Can you count the words? \\
&Refine Prompt
&C.RP&Metacognitive&Heuristic&The learner improves the prompt to get better feedback.&1. Instead of just ..., can you ...? \\
&Correct Answer&C.CA&Metacognitive&Heuristic&The learner identifies and corrects inaccuracies, outdated content, or misleading information produced by the AI.&1. You made a mistake, I think the correct answer is ... \\ \midrule
Reflection &Self-Evaluate&R.SE&Metacognitive&Theoretical&The learner evaluates their used strategies or gained knowledge after task.
&1. I feel confident about ... but still struggle with ...\newline
2. What is the best way for me to ...? \\
&Summary Learning&R.SL&Metacognitive&Theoretical&The learner summarizes what they have learned in their own words.&1. Today I learned... \\
&Plan Next Step&R.PN&Metacognitive&Theoretical&The learner plans the subsequent learning actions, such as reading, reviewing, or practicing.&1. I will try to ... in the next stage. \\ \midrule
-&Acknowledgment&N/A&Non-SRL&Heuristic&The learner acknowledges AI responses but not related to learning process.&1. Thanks! ;
2. Morning.  \\
\bottomrule
\end{tabular}
\end{minipage}
\end{adjustbox}
\end{table}
\restoregeometry

\subsection{Empirical Implementation}
\label{subsec:proof-of-concept}
To evaluate the applicability of the Gen-SRL annotation schema in authentic chatbot-mediated learning settings,  we applied it to a real-world dataset of student–chatbot interactions from a university writing support course. In this section, we describe the data collection, the procedures of translating raw data to SRL actions, and data analysis techniques used to operationalize the Gen-SRL schema and extract meaningful patterns of SRL behaviour to support subsequent analysis.

\subsubsection{Data Collection}
\label{subsubsec:data collection}
Due to the lack of publicly available datasets capturing learner–chatbot interactions in GenAI-powered environments~\cite{2025zixin}, and given that writing tasks naturally elicit diverse SRL behaviors~\cite{graham2000role}, we selected the RECIPE4U dataset~\cite{han-etal-2024-recipe4u-student} as a suitable context for analyzing SRL patterns. This dataset contains complete dialogue logs of 212 university students interacting with a chatbot for English argumentative writing tasks. These dialogues primarily focused on text polishing and writing support. The dataset includes 504 sessions (407 multi-turn), comprising 1,913 student prompts and 2,417 chatbot responses. Each log entry contains a case ID, timestamp, and the corresponding student prompt and chatbot response.

All interactions were collected during a 7-week English-as-a-Foreign-Language (EFL) writing course on the RECIPE platform~\cite{han2023exploring}. The chatbot was powered by a fine-tuned GPT-3.5 model and acted as a writing tutor. To promote engagement, the chatbot initiated each session by asking reflective questions such as “What did you learn today?” All data were collected with participant consent and fully anonymized before public release.

\subsubsection{Translation of Raw Data to SRL Microlevel Action}
To convert raw chatbot logs into analyzable SRL microlevel actions, we performed preprocessing, manual annotation, and inter-annotator agreement assessment, as illustrated below.

\noindent\textbf{Preprocessing }\hspace{0.5em}
To prepare the dataset for annotation and analysis, we first developed an automated preprocessing script to transform raw conversational logs (in JSON format) into structured event records (in CSV format) suitable for subsequent annotation and analysis. Each JSON conversation unit was converted into a single row, retaining case ID, timestamp, and the prompt content. 

We then applied two preprocessing steps:
1) Prompt Segmentation: If a single student prompt contained multiple distinct SRL behaviours, it was split into separate event records, each corresponding to one microlevel SRL action;
2) Length Filtering: Conversations with fewer than 10 meaningful prompts were excluded from the dataset, as such short sequences typically represented incomplete or trivial learning episodes~\cite{bannert2014process}.

\noindent\textbf{Annotation}\hspace{0.5em}
Each student's prompt was manually annotated using the Gen-SRL schema. Specifically, the annotators manually assigned each prompt to a corresponding action code by analyzing its semantic content, intent, and conversational context, referencing predefined microlevel action descriptions and prompt-action dictionary.

The annotation process follows two constraints: (1) Prompts were labeled in the order they appeared in the conversation; 
(2) Each prompt was assigned exactly one microlevel action label;

The annotation process also involved several additional considerations. Firstly, non-SRL prompts labeled as \textit{N/A} were excluded from further modelling. In addition, the dataset included system-initiated scaffolding prompts, such as the chatbot asking "What did you learn today?" at the beginning. Student responses to such prompts were not spontaneous and did not reflect self-regulation behaviours~\cite{li2023analytics,roll2011metacognitive}. Therefore, we annotated these prompts but marked them with an asterisk (e.g., \textit{R.SL*} in Figure~\ref{fig:snippet}), and excluded them from the modelling.

\noindent\textbf{Inter-Annotator Agreement }\hspace{0.5em}
Inter-Annotator Agreement (IAA) is primarily carried out to evaluate consistency among multiple annotators in labeling categorical data. In our study, two trained graduate students independently labeled each conversation based on the Gen-SRL schema. We adopted the nominal level of measurement, since each conversation turn was classified into a specific SRL category based on its content. Consequently, Cohen's Kappa \cite{cohen1960coefficient} was used to measure the level of agreement between two annotators, defined as:

\begin{equation}
\kappa = \frac{P_o - P_e}{1 - P_e}
\end{equation}

where $P_o$ is the observed agreement and $P_e$ is the expected agreement by random chance. Kappa values range from $-1$ (perfect disagreement) to $1$ (perfect agreement), with $0$ indicating agreement at the level of random chance. Higher Kappa values indicate stronger inter-rater reliability. The annotated data were then used for further data analysis.

\subsubsection{Data Analysis}
\label{subsec:data analysis}
In this section, we conducted both frequency analysis and process analysis as a combined approach to describe students' SRL behaviours based on the annotated dataset.

For frequency analysis, we performed descriptive statistics to quantify behaviour distributions across SRL phases and actions. For each SRL phase and action, we computed the total number of occurrences (N), percentage of total actions (\%), number of cases in which the action occurred at least once, also known as case frequency (CF), percentage of learners who engaged in the action (CF\%), the minimum and maximum number of times each action was performed by an individual learner, the mean number of occurrences per learner (M) and the standard deviation (SD).

To discover and visualize the sequential structure of SRL behaviour, we used the Disco tool~\cite{disco}, which applies the Fuzzy Miner algorithm~\cite{fuzzyminer} to extract dominant behavioural paths while filtering out infrequent or noisy transitions, making it especially suitable for educational settings. We generated both microlevel and macrolevel process models based on the annotated dataset. The microlevel model was discovered using the annotated SRL actions, while the macrolevel model was derived by grouping actions according to their corresponding SRL phases.

The key attributes for PM were configured as follows:
(1) Case: the case ID (as introduced in Section~\ref{subsubsec:data collection});
(2) Activity: either the SRL microlevel action or macrolevel phase, depending on the model;
(3) Timestamp: the timestamp of each action;

To improve the interpretability of discovered process models,  we retained all activities (i.e., 100\% activity detail) and adjusted the path detail to hide lower-frequency transitions (all process models are labeled with their respective detail settings). These settings help avoid overly complex \textit{Spaghetti} models and improve model clarity \cite{vanderAalst2016}. For completeness, we also generate the full transition frequency matrix to capture all behavioural transitions, including those omitted by these filtering settings.

\section{Results}
\label{sec:results}
\subsection{Annotation Overview}
\begin{figure}[htbp]
\centering
\includegraphics[width=\textwidth]{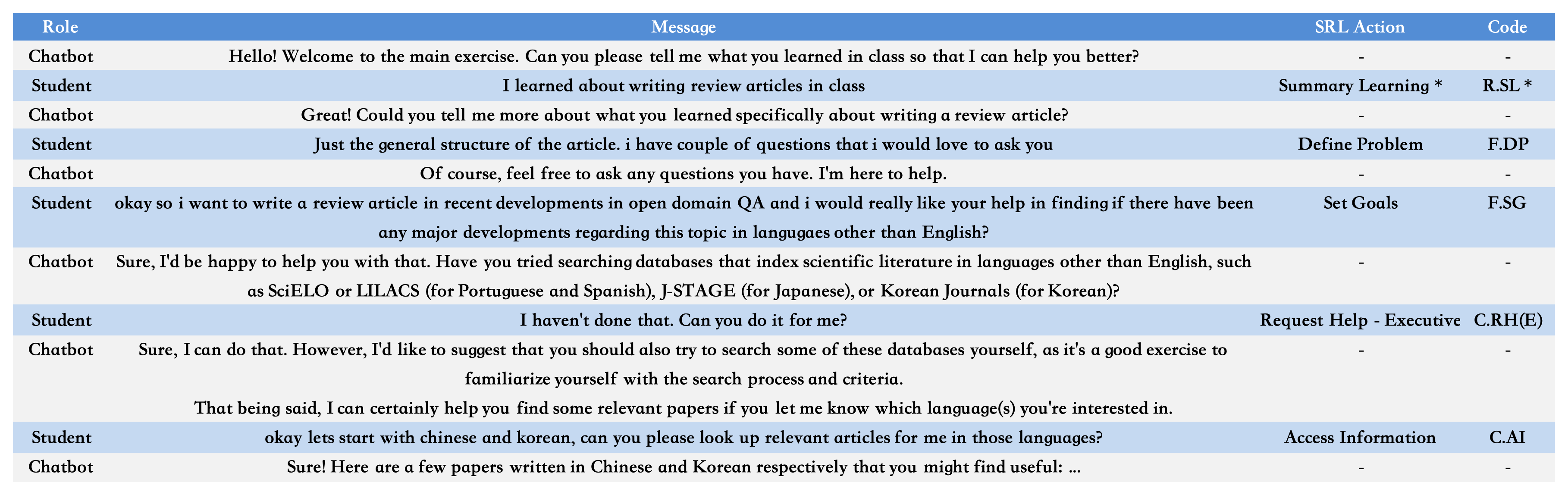}
\caption{Annotated snippet of a student–chatbot conversation showing SRL microlevel actions based on the Gen-SRL annotation schema.}
\label{fig:snippet}
\end{figure}

We manually annotated student–chatbot interactions from the RECIPE4U dataset aforementioned in Section \ref{subsubsec:data collection}. After preprocessing, 1058 utterances from 42 students remained. All utterances were successfully annotated to specific SRL microlevel actions using the Gen-SRL annotation schema. Note that 23 non-SRL actions and 141 system-initiated scaffolding actions were marked with an asterisk and excluded from subsequent analysis. As a result, the final dataset contained 894 valid utterances, with an average of 21.3 per student. In particular, all 16 SRL microlevel actions defined in our schema were represented in the annotated data. A snippet of a conversation that exhibits this is shown in Figure~\ref{fig:snippet}.


\subsection{Inter-Annotator Agreement and Reliability}
After completing the annotation process, we evaluated the reliability between annotators for the RECIPE4U dataset using Cohen's Kappa ($\kappa$), a statistical measure of agreement for nominal data. Two trained graduate students independently annotated all student–ChatGPT conversation prompts based on the Gen-SRL annotation schema. The annotation yielded strong agreement with $\kappa = 0.851$.

To evaluate the robustness and statistical accuracy of this agreement value, we employed bootstrap resampling~\cite{efron1992bootstrap}. Specifically, we generated 1000 bootstrap samples by resampling the annotated data with replacement. For each sample, we recalculated Cohen's Kappa to obtain an empirical distribution of $\kappa$ values.

Based on this distribution, we estimated the 95\% confidence interval for the RECIPE4U dataset to be (0.823, 0.871). This interval suggests that the true underlying agreement value is likely to fall within this range with 95\% confidence, which reinforces the reliability of the annotation procedure.

\subsection{SRL Patterns in Chatbot-Powered Learning Environments}
To answer \textbf{RQ2}, this section presents descriptive findings on students' SRL behaviors in chatbot-supported learning environments. We begin with a frequency-based overview, followed by an analysis of the macro- and micro-level process models to illustrate the sequential SRL patterns. These results serve as the foundation for further discussion in Section~\ref{sec:discussion}.

\subsubsection{Frequency Analysis}
\label{subsubsec:frequcy analysis}
\begin{table}[htbp]
\centering
\caption{Frequencies and descriptive measures of SRL phases and corresponding microlevel actions based on annotated chatbot interactions. All metric definitions are provided in Section~\ref{subsec:data analysis}.}
\label{tab:srl_micro_stats}
\begin{tabularx}{\textwidth}{>{\raggedright\arraybackslash}p{3.2cm} >{\raggedright\arraybackslash}p{4.5cm} >{\raggedright\arraybackslash}p{1.8cm} >{\raggedright\arraybackslash}p{0.9cm} >{\raggedright\arraybackslash}p{1.0cm} >
{\raggedright\arraybackslash}p{0.9cm} >
{\raggedright\arraybackslash}p{1.0cm} >{\raggedright\arraybackslash}p{0.9cm} >{\raggedright\arraybackslash}p{0.9cm} >{\raggedright\arraybackslash}p{0.9cm} >{\raggedright\arraybackslash}p{0.8cm}}
\toprule
\textbf{Macrolevel Phase} & \textbf{Microlevel Action} & \textbf{Code} & \textbf{N} & \textbf{\%} & \textbf{CF} & \textbf{CF (\%)} & \textbf{Min} & \textbf{Max} & \textbf{M} & \textbf{SD} \\
\midrule

\textbf{Forethought} & \textbf{Phase Total} &  & \textbf{48} & \textbf{5.37} & \textbf{16} & \textbf{38.09} & \textbf{0} & \textbf{8} & \textbf{1.14} & \textbf{1.77} \\
\quad & Set Goals & F.SG & 21 & 2.35 & 13 & 30.95 & 0 & 4 & 0.50 & 0.91 \\
\quad & Define Problem & F.DP & 27 & 3.02 & 13 & 30.95 & 0 & 6 & 0.64 & 1.25 \\

\midrule
\textbf{Monitoring} & \textbf{Phase Total} &  & \textbf{95} & \textbf{10.63} & \textbf{28} & \textbf{66.67} & \textbf{0} & \textbf{17} & \textbf{2.26} & \textbf{3.15} \\
\quad & Check Understanding - Basic & M.CU(B) & 79 & 8.84 & 26 & 61.90 & 0 & 14 & 1.88 & 2.79 \\
\quad & Check Understanding - Deep & M.CU(D) & 16 & 1.79 & 11 & 26.19 & 0 & 3 & 0.38 & 0.72 \\

\midrule
\textbf{Control} & \textbf{Phase Total} &  & \textbf{736} & \textbf{82.32} & \textbf{41} & \textbf{97.62} & \textbf{0} & \textbf{66} & \textbf{17.52} & \textbf{14.33} \\
\quad & Seek Clarification-Basic & C.SC(B) & 91 & 10.18 & 30 & 71.43 & 0 & 14 & 2.17 & 2.83 \\
\quad & Seek Clarification - Deep & C.SC(D) & 43 & 4.81 & 18 & 42.86 & 0 & 9 & 1.02 & 1.88 \\
\quad & Request Help - Instrumental & C.RH(I) & 143 & 16.00 & 28 & 66.67 & 0 & 14 & 3.45 & 5.05 \\
\quad & Request Help - Executive & C.RH(E) & 145 & 16.22 & 33 & 78.57 & 0 & 15 & 3.40 & 4.76 \\
\quad & Request Help - Check & C.RH(C) & 184 & 20.58 & 33 & 78.57 & 0 & 24 & 4.38 & 6.88 \\
\quad & Access Information & C.AI & 97 & 10.85 & 22 & 52.38 & 0 & 14 & 2.31 & 4.27 \\
\quad & Refine Prompt & C.RP & 17 & 1.90 & 15 & 35.71 & 0 & 2 & 0.40 & 0.58 \\
\quad & Correct Answer & C.CA & 16 & 1.79 & 10 & 23.81 & 0 & 4 & 0.38 & 0.84 \\

\midrule
\textbf{Reflection} & \textbf{Phase Total} &  & \textbf{15} & \textbf{1.68} & \textbf{11} & \textbf{26.19} & \textbf{0} & \textbf{4} & \textbf{0.36} & \textbf{0.78} \\
\quad & Self-Evaluate & R.SE & 4 & 0.45 & 4 & 9.52 & 0 & 1 & 0.10 & 0.29 \\
\quad & Plan Next Step & R.PN & 8 & 0.89 & 6 & 14.29 & 0 & 3 & 0.19 & 0.55 \\
\quad & Summary Learning & R.SL & 3 & 0.34 & 3 & 7.14 & 0 & 1 & 0.07 & 0.26 \\

\bottomrule
\end{tabularx}
\end{table}
 
Table~\ref{tab:srl_micro_stats} presents the descriptive statistics of all annotated SRL phases and actions in the RECIPE4U dataset, covering 894 actions from 42 learners. These results were generated based on the frequency analysis procedures described in Section~\ref{subsec:data analysis}. Here, we conducted a descriptive analysis to examine the distribution of students' SRL behaviours across different phases and actions (assess \textbf{H1}).

At the macro level, the \textit{Control} phase  clearly dominated the SRL process, both in absolute frequency (N = 736, 82. 32\%) and in case frequency (CF = 41, 97.62\%). This means that more than 82\% of all annotated prompts reflected control-phase behaviors, and nearly all learners engaged in at least one control-phase action. This greatly exceeded the engagement observed in \textit{Monitoring} (N = 95, 10.63\%, CF = 28, 66.67\%), \textit{Forethought} (N = 48, 5.37\%, CF = 16, 38.09\%), and \textit{Reflection} (N = 15, 1.68\%, CF = 11, 26.19\%). 

Within the \textit{Control} phase, help-seeking behaviours were the most prominent, i.e., \textit{Request Help – Check} (N = 184, 20.58\%, CF = 33, 78.57\%), \textit{Request Help – Instrumental} (N = 143, 16.00\%, CF = 28, 66.67\%), and \textit{Request Help – Executive} (N = 145, 16.22\%, CF = 33, 78.57\%). These behaviours also showed high variability between learners (SD = 6.88, 5.05, and 4.76, respectively). In contrast, metacognitive actions in \textit{Control} phase such as \textit{Correct Answer} (N = 16, 1.79\%, CF = 10, 23.81\%, SD = 0.84) and \textit{Refine Prompt} (N = 17, 1.90\%, CF = 15, 35.71\%, SD = 0.58) were used much less frequently and showed lower individual variation. Compared to \textit{Control} phase, every microlevel action in \textit{Forethought} and \textit{Reflection} phases exhibited even lower levels of engagement. For example, \textit{Set Goals} (N = 21, 2.34\%, CF = 13, 30.95\%, SD = 0.91) and \textit{Summary Learning} (N = 3, 0.34\%, CF = 3, 7.14\%, SD = 0.26) were rarely used and showed consistently low usage among learners. We further computed the proportion of metacognitive actions as defined in our Gen-SRL schema (Table~\ref{tab:srl-framework}) and found that only 26.17\% of the 894 annotated SRL actions were metacognitive, while 73.83\% were cognitive.
 

\subsubsection{Process Analysis}
\label{subsubsec:process analysis}
While the frequency analysis provides a general overview of learners' engagement with distinct SRL behaviours (see Section~\ref{tab:srl_micro_stats}), it does not capture the sequential structure \cite{saint2020combining}. To complement this, we conducted a process analysis based on the discovered macrolevel and microlevel process model to further reveal learners' self-regulation patterns in the chatbot context (access \textbf{H2}).

\begin{figure}[htbp]
  \centering
  \includegraphics[width=0.85\linewidth]{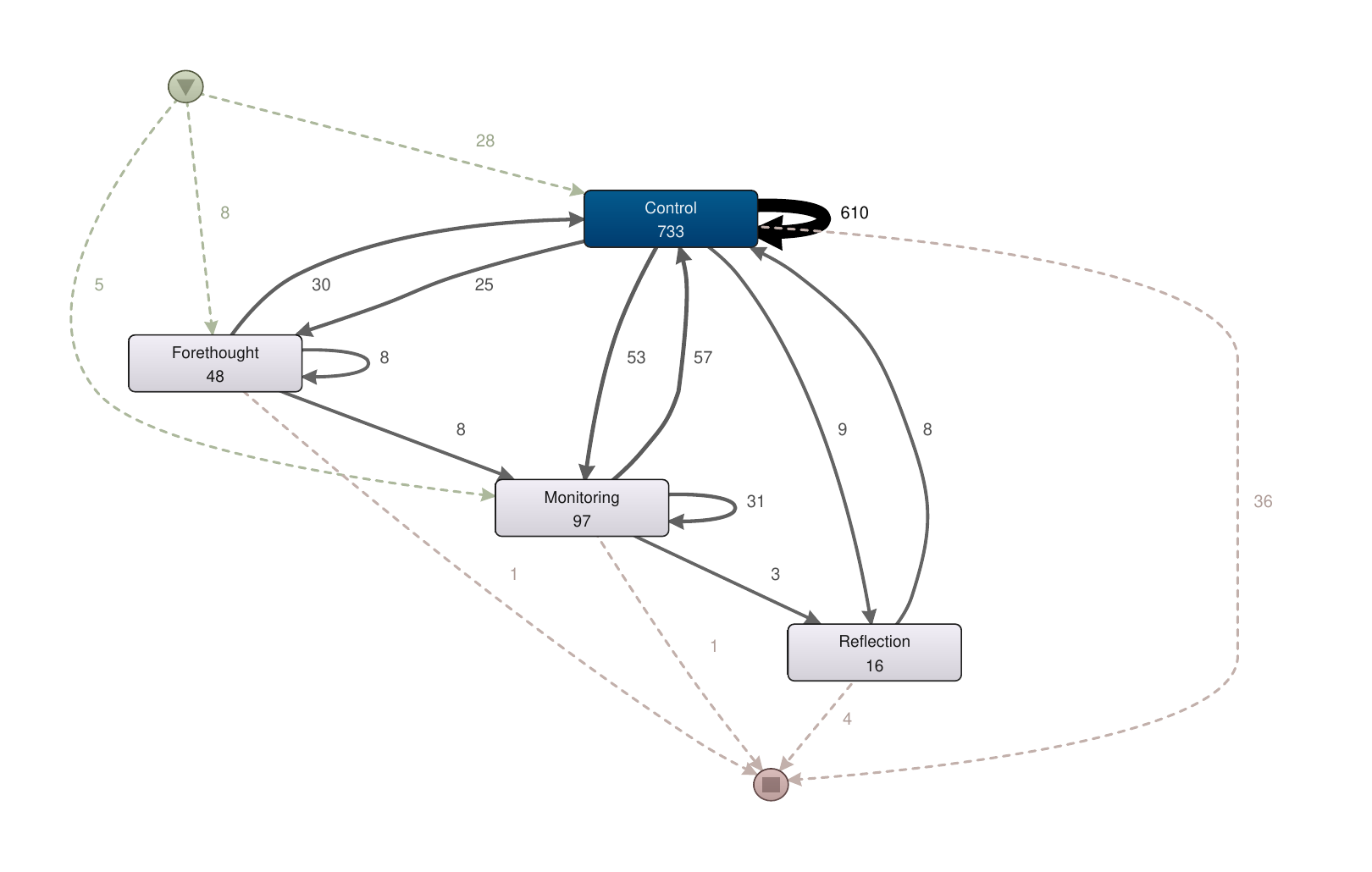}
    \caption{Macrolevel SRL process model generated using Disco with activity and path details set to 100\% and 60\%, respectively. Rectangles represent SRL phases; green/red nodes indicate start/end events. Node shade and edge thickness reflect frequency and transition counts, respectively. All values are absolute frequencies. The full transition frequency matrix is provided in Appendix B for reference.}
  \label{fig:macro_recipe4u}
\end{figure}

\noindent\textbf{Macrolevel }\hspace{0.5em}
Figure~\ref{fig:macro_recipe4u} presents the macrolevel process model discovered in Section \ref{subsec:data analysis}, capturing transitions across four phases. The \textit{Control} phase shows the highest engagement, indicated by its highest frequency (N = 733) and the darkest node color. Notably, most SRL sequences started from \textit{Control} (n = 28, where n denotes the number of paths) and ended in \textit{Control} (n = 36), with fewer starting in \textit{Forethought} (n = 8) or ending in \textit{Reflection} (n = 4). Frequent bidirectional transitions were also observed between \textit{Control} and other phases, particularly \textit{Monitoring} (n = 53) and \textit{Forethought} (n = 25). Furthermore, all phases exhibited internal self-loops, most prominently in \textit{Control} (n = 610). These patterns were not apparent from frequency statistics but became evident through process visualization.

\begin{figure}[htbp]
  \centering
  \includegraphics[width=\linewidth]{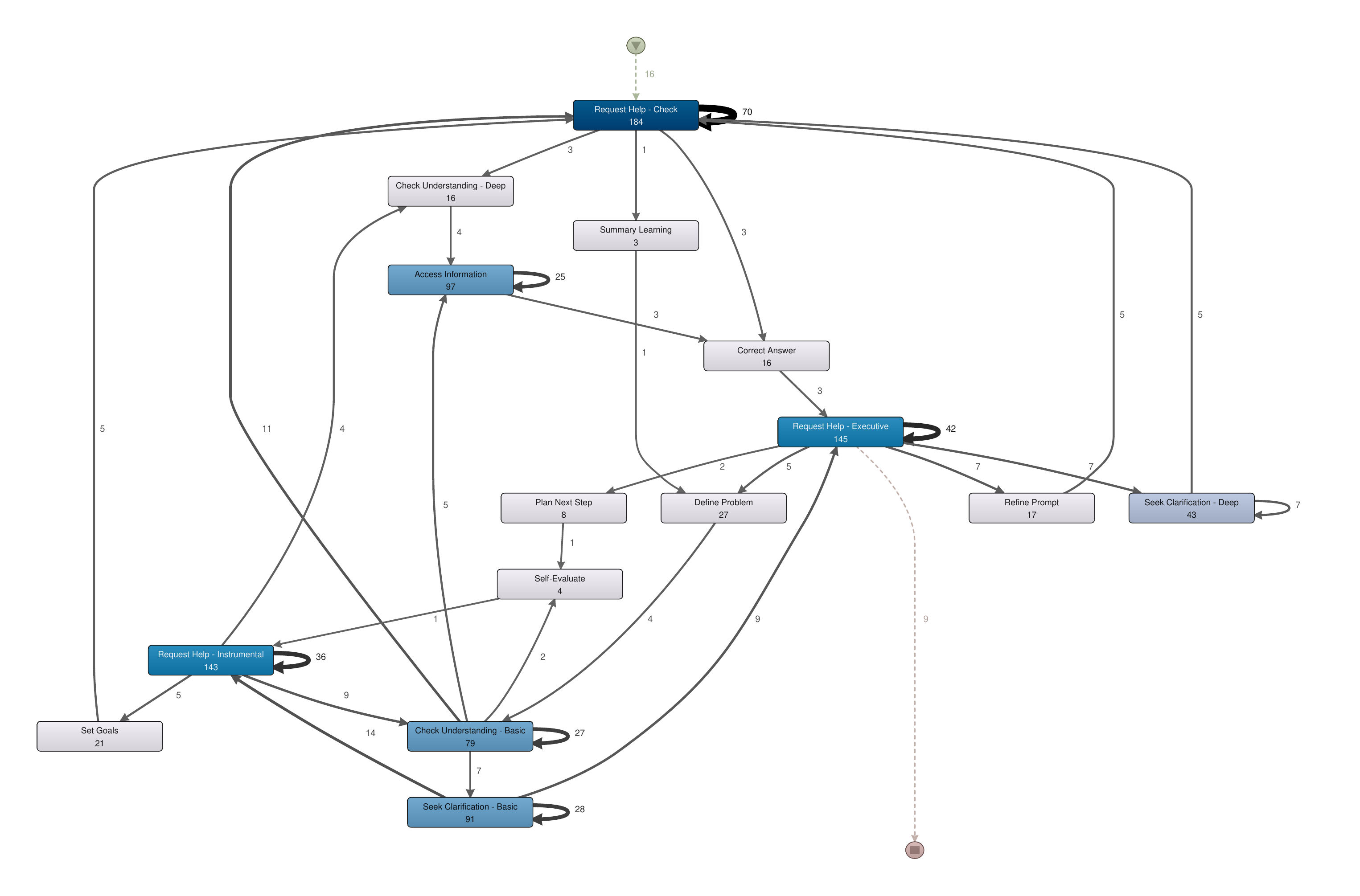}
    \caption{Microlevel SRL process model generated using Disco with activity and path details set to 100\% and 17.8\%, respectively. Visual elements follow Figure~\ref{fig:macro_recipe4u}; all values represent absolute frequencies. The full transition frequency matrix is provided in Appendix B for reference.}
  \label{fig:recipe4u_micro_case}
\end{figure}

\noindent\textbf{Microlevel }\hspace{0.5em}
 Building on the macrolevel findings, we applied a microlevel lens to analyse learners' action-level transitions, as visualized in the microlevel process model shown in Figure~\ref{fig:recipe4u_micro_case}.

This model clarifies the behavioral composition underlying the macrolevel transitions. For instance, our macrolevel analysis showed that students often began and ended their SRL cycles through \textit{Control} phase, but the microlevel model revealed that learners frequently started with \textit{Request Help – Check} and finished with \textit{Request Help – Executive}. Similarly, the dominant self-loop in \textit{Control} is primarily concentrated on specific nodes, particularly \textit{Request Help-Check} and \textit{Request Help-Executive}, rather than all control-phase behaviours. Note that the microlevel model in Figure~\ref{fig:recipe4u_micro_case} is not used to directly validate our hypotheses, but to explain macro patterns and guide specific interventions.

\section{Discussion}
\label{sec:discussion}
\subsection{Imbalanced SRL Behaviours in Chatbot-Powered Learning}
\label{subsec:discussion1}

Our frequency analysis in Section \ref{subsubsec:frequcy analysis} reveals an imbalance distribution of SRL behaviours. More than 82\% of all actions and 97\% of learners are concentrated in the \textit{Control} phase, indicating a strong behavioral bias toward in-task regulation. In contrast, engagement with the \textit{Forethought} and \textit{Reflection} phases was minimal. This skewed distribution aligns with Greene et al.~\cite{greene2011}, who highlighted that students tend to rely heavily on a narrow set of in-task strategies while overlooking phases involving metacognitive regulation. Beyond this phase-level distribution, we also observed an internal imbalance within the \textit{Control} phase. Specifically, although \textit{Request Help} behaviours were prevalent, the limited use of metacognitive actions such as \textit{Refine Prompt} and \textit{Correct Answer}, which have been shown to promote high-order thinking and subsequent self-regulation behaviours~\cite{walter2024embracing,lee2024empowering}. This indicates a tendency to engage with AI feedback in a surface-level manner, where students frequently seek help but rarely critically regulate or evaluate feedback, pointing to a reliance on AI as an external problem solver rather than a resource to scaffold deeper self-regulation. It highlights the need to design AI systems as scaffolds that promote deeper self-regulation behaviours.

\begin{figure}[htbp]
  \centering
  \includegraphics[width=0.7\textwidth]{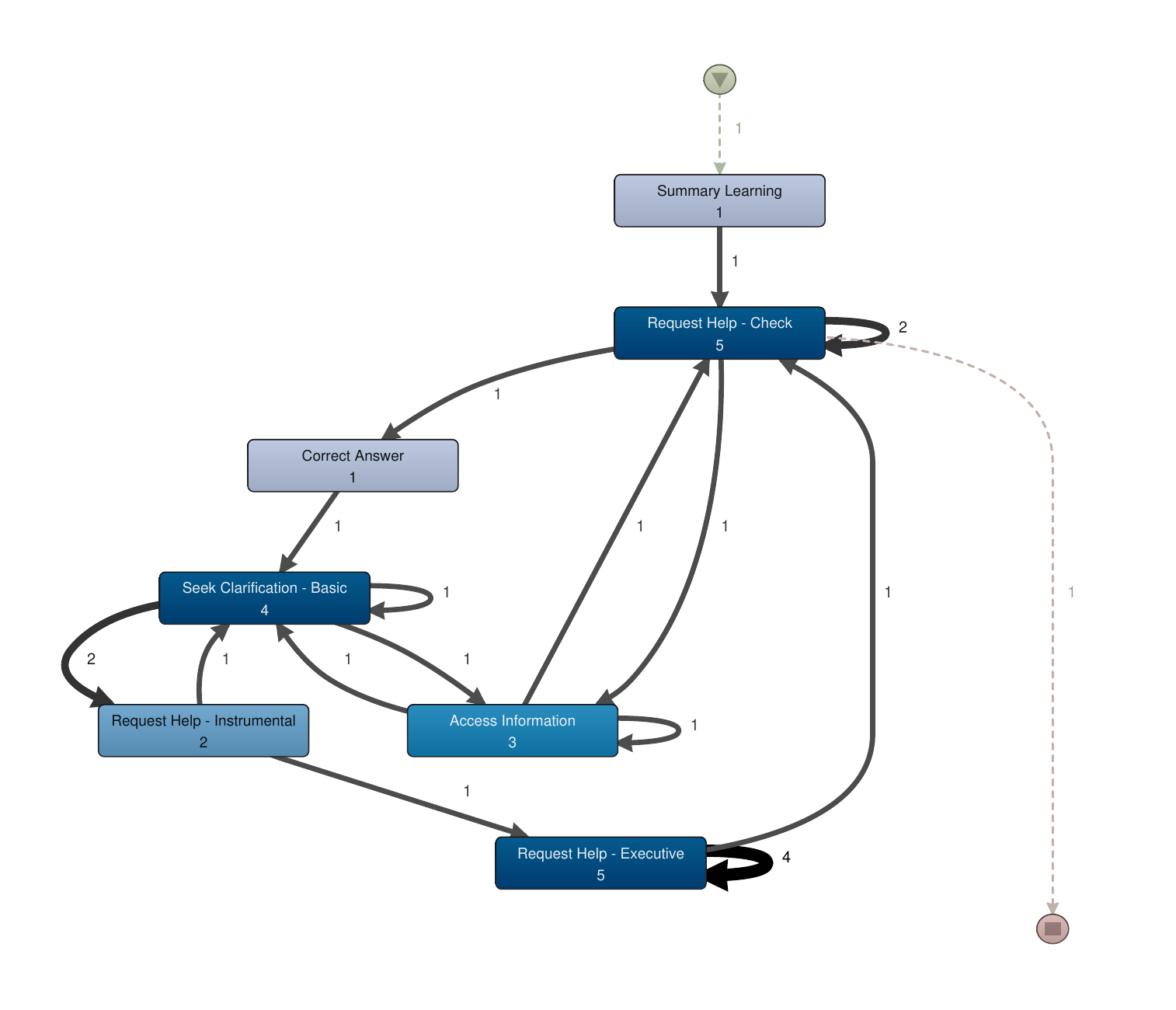}
  \caption{Microlevel process model of case A: Balanced use of Control-phase strategies (activity and path detail: 100\%).}
  \label{fig:caseA}
\end{figure}

\begin{figure}[htbp]
  \centering
  \includegraphics[width=0.7\textwidth]{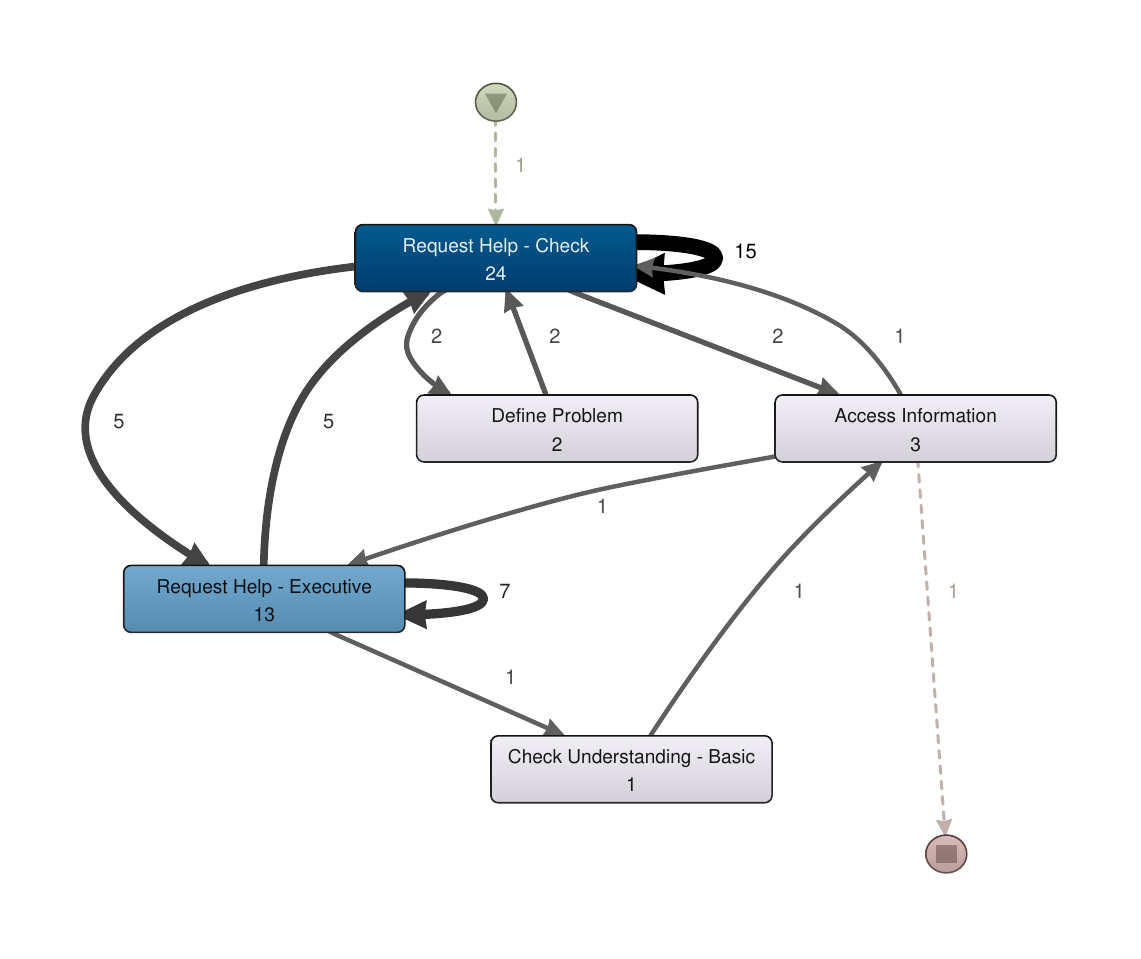}
  \caption{Microlevel process model of case B: Repetitive help-seeking behaviour in the Control phase (activity and path detail: 100\%).}
  \label{fig:caseB}
\end{figure}

In addition to general imbalance, we argue that SRL behaviours are also imbalanced at the individual level. Our standard deviation results showed substantial individual variability in the use of help seeking strategies. That is, although \textit{Request Help} behaviours were common, learners varied widely in how often they relied on them. To illustrate the variability in learners' self-regulation strategies, we purposefully selected two contrasting cases from the 42 students that exemplify distinct patterns of control-phase engagement (Figures~\ref{fig:caseA} and~\ref{fig:caseB}). Case A a diverse and balanced use of control-phase actions, whereas Case B heavily relied on repeated \textit{Request Help–Check}, with nearly 4.8 times more than case A. This contrast exemplifies the high standard deviation observed in these behaviours, manifesting the imbalanced nature of individual behavioural patterns in SRL. Meanwhile, behaviours in \textit{Forethought} and \textit{Reflection} showed both low frequency and minimal variability, indicating not only underuse, but a near-consensus absence across learners. These findings suggest that while students differ greatly in the way they use certain strategies (especially within the \textit{Control} phase), there is a shared tendency to underutilize phases requiring metacognitive engagement. Taken together, these findings challenge \textbf{H1} and indicate that SRL behaviours exhibit both imbalance at the group level and high variability at the individual level, underscoring the need for personalized and adaptive SRL scaffolding in GenAI-mediated learning environments, particularly scaffolds that strengthen learners' engagement in metacognitive phases \cite{sobocinski2017exploring}.

To better understand the cause of the observed structural imbalance, we turn to a representative pattern identified in our microlevel process analysis (Section~\ref{subsubsec:process analysis}). Many students began their conversations with prompts like "Can you tell me what is wrong with this paragraph?", without first articulating the problem themselves. In these cases, the chatbot directly pointed out issues such as unclear logic or grammatical errors, then learners immediately accepted them without critical thinking. This pattern exemplifies a “you tell me” learning habit, where learners rely on external feedback instead of engaging in internal regulation. This behaviour reflects a disproportionate focus on control-phase actions, leading to the observed behavioural imbalance. Based on the example, we argue that a key driver of this pattern is metacognitive laziness\cite{fan2025beware}—a tendency to outsource higher-order regulatory tasks, such as planning and evaluation, to external tools such as chatbot\cite{risko2016}. Urban et al.~\cite{urban2024chatgpt} further emphasize that the fluent and instant responses provided by AI can lower students' perceived task difficulty, but at the cost of reduced cognitive effort and impaired development of metacognitive skills~\cite{molenaar2022a}.

While metacognitive laziness captures a cognitive explanation for structural imbalance, motivational factors may also contribute to the limited engagement of learners in metacognitive phases. Previous research has shown that when interacting with human teachers or peers, students experience a stronger sense of social responsibility and dialogue, which increases their likelihood of expressing planning and evaluative thoughts~\cite{kim2022perceived,chen2025unpacking}. In contrast, when interacting with chatbot, some students treat it as a problem-solving tool rather than a learning partner, which may suppress their motivation to set goals or reflect during the interaction, contributing to the underuse of \textit{Forethought} and \textit{Reflection} behaviours.

To address the observed imbalance, educational interventions must directly target its underlying causes, namely, the tendency of students to offload metacognitive regulation and their limited motivation to engage in planning and reflection. Given that human–AI collaboration is likely to become a central feature of future learning \cite{luckin2016intelligence}, it is critical to design interventions that reactivate learners' metacognitive agency and reframe their relationship with chatbot. An effective strategy is to embed dynamic task-level reflection mechanisms into the learning process \cite{fan2025beware}. Instead of providing direct answers, the chatbot can use a Socratic approach~\cite{paul2019thinker} by asking guiding questions that prompt student reasoning. For instance, when students ask the chatbot to check problems in their writing, rather than directly pointing out flaws, the AI might respond with a prompt like "What do you think could be improved here? Maybe you could start by thinking about any grammar issues." This encourages students to engage in planning, monitoring, and evaluation in real time, while discouraging direct help-seeking. Crucially, the observed individual variability suggests that such prompts should be adaptively tailored to learners' prior behaviors and self-regulation patterns, rather than delivered uniformly. To address motivational problems, educators can design the chatbot to adopt specific roles (e.g., tutor, peer, or reviewer), which may enhance learners' sense of social presence and stimulate more active planning and evaluation behaviours~\cite{jarvela2021,li2025facilitating}. At the same time, learners need support to develop the ability to ethically and effectively delegate tasks to AI. As suggested by Molenaar's HHAIR model~\cite{molenaar2022a}, this division of labor should be dynamic and intentional, helping students strengthen rather than replace their metacognitive skills.
 
\subsection{Deviations from the Sequential SRL Pattern in Chatbot-Powered Learning}
\label{subsec:discussion2}
As described in Section \ref{subsec:srl_patterns}, classical SRL models conceptualize self-regulation as a sequential process. Building on this theoretical expectation, this section access the validity of \textbf{H2} by examining deviations from this cycle, based on learners' interaction patterns derived from process analysis (see Section~\ref{subsubsec:process analysis}).

The advantage of PM is that it can visualize how sequences of regulated learning unfold over time \cite{jarvela2021}. Compared to the sequential transitions assumed in the theoretical SRL models, our macrolevel model in Figure~\ref{fig:macro_recipe4u} shows dense and interconnected connections between phases. There is a strong trend where most SRL processes did not begin with the \textit{Forethought} phase, as suggested by theory (i.e., the expected \textit{Start} → \textit{Planning} transition in Figure~\ref{fig:traditional_srl_model}), but instead started directly from the \textit{Control} phase. As we discussed before, learners often began by asking the chatbot to check issues, then used its feedback to better understand their situation and further define problems. This indicates that planning tends to occur retrospectively after task execution during the student-chabot interactions, rather than as a pre-task strategy emphasized in theoretical models. Similarly, most of the students did not end their learning processes with the \textit{Reflection} phase (i.e., the expected \textit{Reflection} → \textit{End} transition in Figure~\ref{fig:traditional_srl_model}). This pattern of "no planning start" and "no reflection end" suggests that the expected phase order is disrupted at both start and end points, illustrating a notable disordered regulation structure unique to GenAI-mediated learning contexts.

We also observed some other nonsequential patterns between SRL phases, including phase skipping and unexpected backward transitions. Phase skipping refers to learners bypassing one or more expected phases in the self-regulated learning cycle, rather than progressing through them sequentially. For example, some learners moved directly from \textit{Control} to \textit{Forethought} without engaging in \textit{Reflection} in Figure~\ref{fig:macro_recipe4u}. This behaviour is a key indicator of disordered SRL, as it weakens the traditional feedback-driven loop. This pattern may result from two reasons: (1) learners skip the \textit{Reflection} phase entirely and initiate a new SRL cycle; (2) learners often ask chatbot to perform an initial check before engaging in planning themselves. Unexpected backward transitions can be viewed as a variant of phase skipping, where learners bypass the next two phases in the cycle and loop back to a previous one. For example, the transition from \textit{Reflection} to \textit{Control} suggests that learners in GenAI-supported environments tend to internalize AI feedback as a form of learning guidance, prompting immediate rapid behavioural adjustments. In contrast, the theoretically expected transitions from \textit{Reflection} to \textit{Forethought}, which mark the beginning of a new SRL cycle, were entirely absent in our discovered model. This missing link disrupts the sequential structure assumed in traditional SRL models.

As discussed in Section~\ref{subsubsec:process analysis}, we observed strong self-loops within individual phases in Figure \ref{fig:macro_recipe4u}, particularly in the \textit{Control} phase, suggesting that learners were often locked in a single-phase regulation without progressing to the next phase. Our microlevel model further revealed that these macrolevel self-loops were not merely repetitions of the same action, but consisted of multiple interrelated microlevel actions (e.g., \textit{Request Help–Check} $\rightarrow$ \textit{Correct Answer} $\rightarrow$ \textit{Request Help–Executive}), forming a local loop that remained within the same phase. These extensive self-loops differ from the previously observed nonsequential pattern and reflect another form of disordered SRL, namely dephasing, where learners remain confined within a single phase instead of progressing sequentially through phases as assumed in the sequential SRL model~\cite{jarvela2019capturing}. Taken together, these patterns contradict \textbf{H2} and align with the earlier work of Järvelä et al. \cite{jarvela2021}, who emphasized the complex and dynamic nature of SRL in authentic learning environments. 

Interestingly, the imbalanced distribution identified in Section \ref{subsec:discussion1}, particularly the behavioural underuse of \textit{Forethought} and \textit{Reflection}, appears to contribute to the observed SRL patterns. That is, the absence of certain phases constrains the formation of complete SRL cycles and may lead to disordered SRL sequences, such as frequent phase skipping. This suggests that the aforementioned causes, such as metacognitive laziness, also help explain the disruption of the overall regulation process. However, we must also consider the influence of the chatbot tool itself in shaping students' SRL behaviours. The real-time nature of chatbot feedback encourages students to engage in learn-through-doing rather than pre-task planning or post-task reflection. As a result, learners may not pause to reflect, but instead continuously process chatbot responses and solve problems, leading to a task-driven, reactive learning environment.

Our results suggest that existing SRL theories may be insufficient to fully capture the chatbot-mediated learning processes, underscoring the need for future research to extend SRL models to better adapt GenAI-supported contexts \cite{xu2025enhancing}. Furthermore, considering the nonsequentail nature of learners' self-regulation behaviours, traditional fixed SRL scaffolding strategies may no longer be effective in chatbot-mediated environments~\cite{lin2023chat}. Many existing SRL-enhanced chatbot designs strictly follow classical sequential SRL theories, where students are expected to reflect and then plan before starting a new task~\cite{ji2024mobile,maldonado2022miranda}. However, our findings indicate that these strategies may conflict with the actual SRL behaviours of the learners. If we attempt to rigidly embed a classical SRL cycle in the chatbot, it may disrupt learners' natural learning flow, potentially leading to cognitive overload or reduced autonomy. Darvishi supports this concern, arguing that combining conventional SRL strategies with GenAI tools without adaptation may be ineffective in supporting both task performance and metacognitive engagement~\cite{darvishi2024impact}. Thus, we argue for a more flexible and context-aware SRL-enhanced scaffolding approach that responds to the situational behaviours of learners \cite{munshi2022analyzing,li2025facilitating}. For example, based on our microlevel process model as illustrated in Figure~\ref{fig:recipe4u_micro_case}, one promising direction is to identify specific precursor actions that frequently precede metacognitive behaviours and use them as scaffold entry points, enabling more responsive regulation support.
 
\section{Limitations and Future Work}
\label{sec:limitations and future work}
This study has several limitations that suggest directions for future research. The current dataset contains only single timestamped interaction events and lacks start–end duration data, making it impossible to estimate the temporal intensity of SRL behaviours (e.g., the time on task). However, not all SRL actions are created equal in depth and cognitive effort. For example, two students may both perform \textit{Set Goals}, but one does so in a few seconds, while another takes a full minute to formulate a thoughtful goal. Such differences are currently invisible, but could be revealed through temporal indicators. Future work should incorporate event duration to better assess the depth and persistence of learners' engagement during self-regulation. In addition, current annotations were conducted manually, future research could explore automatic tagging of SRL behaviours by enhancing the prompt–action dictionary in the Gen-SRL annotation schema, allowing machine learning models to more accurately infer SRL actions from chatbot dialogues.

Second, our analysis focused on a single classroom dataset focused on an EFL writing task, which limits generalizability. Previous studies have highlighted that SRL behaviours often vary across academic domains and task types \cite{ainley2006measuring,bannert2014}, suggesting that future studies should explore diverse academic settings. In addition to cross-domain comparisons, future studies could address this by contrasting students with high and low academic performance to examine how their SRL behaviour patterns diverge. Such cross-domain and performance-based analyses would support the development of more targeted and context-sensitive SRL scaffolds in chatbot-mediated environments.

Furthermore, the study used Fuzzy Miner for better visualization. Although suitable for highlighting frequent behaviours, this algorithm may suppress lower-frequency but educationally significant transitions \cite{van2012process1}. Future work should compare alternative mining approaches (e.g., inductive mining, heuristic mining), or combine multiple techniques to ensure robustness and interpretability.

Finally, some key metacognitive behaviours, such as setting goals or self-evaluation, may have occurred implicitly and were not explicitly reflected in student-chatbot interactions. This issue echoes a well-recognized challenge in SRL research: many regulation processes are inherently internal and difficult to detect by surface-level traces \cite{bannert2014}. For instance, learners may have evaluated the chatbot's feedback or planned their next steps internally without verbalizing these thoughts. In this context, relying solely on observable dialogue data may limit the accuracy of the analysis. To address this, future research should incorporate complementary methods such as think-aloud protocols to measure the real-time psychological processes of learners \cite{azevedo2017understanding,vandevelde2015using}. In addition, as recent multimodal learning analytics research shows, incorporating sources such as screen recordings, think-aloud data, eye-tracking, and physiological signals can reveal hidden regulatory processes and enhance ecological validity \cite{jarvela2024triggers,cloude2022system}.

\section{Conclusion}
\label{sec:conclusion}
In conclusion, this study proposes Gen-SRL, a hierarchical annotation schema grounded in recognised SRL theory and tailored to capture SRL behaviours in chatbot-mediated interactions. Coupled with frequency analysis and PM techniques, we provide the first systematic approach to process and analyse SRL behaviours in GenAI-supported learning environments. Our analysis revealed two key findings. First, the distribution of SRL behaviours was imbalanced and varied widely across learners, with most actions concentrated in task execution and limited engagement in planning and reflection. Second, the expected SRL flow was often nonsequential, showing patterns such as beginning without planning, ending without reflection, phase skipping, and self-loops. Further analysis showed that the patterns may be due to learners' metacognitive laziness, limited social presence, and the task-driven nature of chatbot. Based on these findings, we emphasize that existing SRL theories cannot fully capture dynamic regulation patterns in GenAI-supported environments. To foster deeper metacognitive engagement, we recommend embedding dynamic, task-level reflection mechanisms such as Socratic prompts. In addition, chatbot should shift from task responders to intelligent regulatory partners, flexibly adopting roles such as tutor, peer, or reviewer. Lastly, SRL scaffolds should be personalized and context-aware. This study advances understanding of SRL patterns in GenAI-mediated learning and offers practical guidance for building educational chatbots that foster self-regulated learning in adaptive and personalized ways.

\section*{Acknowledgements} The author gratefully acknowledges the academic supervision and support provided by Abel Armas Cervantes and Antonette Mendoza throughout this research. Special thanks also go to Ren Ding, from the Master of Software Engineering program at the University of Melbourne, for his contribution to the annotation process.

\section*{Data Availability Statement} The data that support the findings of this study are openly available at \url{https://github.com/lvyl9909/SRL-Research-Project}, including the complete annotated dataset and preprocessing scripts.

\section*{Ethics Statement} This study is a secondary analysis of the anonymized RECIPE4U dataset, which was originally collected by its authors under IRB approval. We obtained permission and submitted a usage consent form. No new data involving human participants was collected. All analyses complied with ethical research standards.

\section*{Declaration of Conflicting Interest} The author declared no potential conflicts of interest with respect to the research, authorship, and/or publication of this article.

\begin{CJK}{UTF8}{mj}
\bibliographystyle{plain}
\bibliography{main}

\begin{thebibliography}{10}

\bibitem{ainley2006measuring}
Mary Ainley and Lyn Patrick.
\newblock Measuring self-regulated learning processes through tracking patterns of student interaction with achievement activities.
\newblock {\em Educational Psychology Review}, 18:267--286, 2006.

\bibitem{araka2020research}
Eric Araka, Elizaphan Maina, Rhoda Gitonga, and Robert Oboko.
\newblock Research trends in measurement and intervention tools for self-regulated learning for e-learning environments—systematic review (2008--2018).
\newblock {\em Research and Practice in Technology Enhanced Learning}, 15:1--21, 2020.

\bibitem{armas2024dusting}
Abel Armas-Cervantes, Ehsan Abedin, and Farbod Taymouri.
\newblock Dusting for fingerprints: Tracking online student engagement.
\newblock {\em Computers and Education: Artificial Intelligence}, 6:100232, 2024.

\bibitem{azevedo2019analyzing}
Roger Azevedo and Dragan Ga{\v{s}}evi{\'c}.
\newblock Analyzing multimodal multichannel data about self-regulated learning with advanced learning technologies: Issues and challenges, 2019.

\bibitem{azevedo2017understanding}
Roger Azevedo, Michelle Taub, and Nicholas~V Mudrick.
\newblock Understanding and reasoning about real-time cognitive, affective, and metacognitive processes to foster self-regulation with advanced learning technologies.
\newblock In {\em Handbook of self-regulation of learning and performance}, pages 254--270. Routledge, 2017.

\bibitem{bannert2014process}
Maria Bannert, Peter Reimann, and Christoph Sonnenberg.
\newblock Process mining techniques for analysing patterns and strategies in students’ self-regulated learning.
\newblock {\em Metacognition and learning}, 9:161--185, 2014.

\bibitem{bannert2014}
Maria Bannert, Peter Reimann, and Christoph Sonnenberg.
\newblock Promoting self-regulated learning through prompts.
\newblock {\em Instructional Science}, 42(2):271--290, 2014.

\bibitem{biwer2021}
Frederik Biwer, Wijaya Wiradhany, Maaike oude Egbrink, Harri{\"e}t Hospers, Sarah Waslander, and Sander Begeer.
\newblock Changes and adaptations: How university students self-regulate their online learning during the covid-19 pandemic.
\newblock {\em Frontiers in Psychology}, 12:642593, 2021.

\bibitem{boekaerts2006far}
Monique Boekaerts and Eduardo Cascallar.
\newblock How far have we moved toward the integration of theory and practice in self-regulation?
\newblock {\em Educational psychology review}, 18:199--210, 2006.

\bibitem{boekaerts2000self}
Monique Boekaerts and Markku Niemivirta.
\newblock Self-regulated learning: Finding a balance between learning goals and ego-protective goals.
\newblock In {\em Handbook of self-regulation}, pages 417--450. Elsevier, 2000.

\bibitem{bushuyev2023bani}
Sergiy Bushuyev, Natalia Bushuyeva, Svetlana Murzabekova, and Maira Khussainova.
\newblock Innovative development of educational systems in the bani environment.
\newblock {\em Scientific Journal of Astana IT University}, pages 104--115, 2023.

\bibitem{butler1995feedback}
Deborah~L Butler and Philip~H Winne.
\newblock Feedback and self-regulated learning: A theoretical synthesis.
\newblock {\em Review of educational research}, 65(3):245--281, 1995.

\bibitem{cerezo2020process}
Rebeca Cerezo, Alejandro Bogar{\'\i}n, Maria Esteban, and Crist{\'o}bal Romero.
\newblock Process mining for self-regulated learning assessment in e-learning.
\newblock {\em Journal of Computing in Higher Education}, 32(1):74--88, 2020.

\bibitem{abel2023uncovering}
Abel~Armas Cervantes, Antonette Mendoza, and Ehsan Abedin.
\newblock Uncovering students' learning pathways: A process mining perspective.
\newblock In {\em 34th Australasian Association for Engineering Education Conference (AAEE2023)}, pages 746--754. Engineers Australia Gold Coast, 2023.

\bibitem{chang2023educational}
Daniel~H Chang, Michael Pin-Chuan Lin, Shiva Hajian, and Quincy~Q Wang.
\newblock Educational design principles of using ai chatbot that supports self-regulated learning in education: Goal setting, feedback, and personalization.
\newblock {\em Sustainability}, 15(17):12921, 2023.

\bibitem{chen2025unpacking}
Angxuan Chen, Mengtong Xiang, Junyi Zhou, Jiyou Jia, Junjie Shang, Xinyu Li, Dragan Ga{\v{s}}evi{\'c}, and Yizhou Fan.
\newblock Unpacking help-seeking process through multimodal learning analytics: A comparative study of chatgpt vs human expert.
\newblock {\em Computers \& Education}, 226:105198, 2025.

\bibitem{2025zixin}
Zixin Chen, Jiachen Wang, Meng Xia, Kento Shigyo, Dingdong Liu, Rong Zhang, and Huamin Qu.
\newblock Stugptviz: A visual analytics approach to understand student-chatgpt interactions.
\newblock {\em IEEE Transactions on Visualization and Computer Graphics}, 31(1):908--918, 2025.

\bibitem{clarizia2018chatbot}
Fabio Clarizia, Francesco Colace, Marco Lombardi, Francesco Pascale, and Domenico Santaniello.
\newblock Chatbot: An education support system for student.
\newblock In {\em Cyberspace Safety and Security: 10th International Symposium, CSS 2018, Amalfi, Italy, October 29--31, 2018, Proceedings 10}, pages 291--302. Springer, 2018.

\bibitem{cleary2012assessing}
Timothy~J Cleary, Gregory~L Callan, and Barry~J Zimmerman.
\newblock Assessing self-regulation as a cyclical, context-specific phenomenon: Overview and analysis of srl microanalytic protocols.
\newblock {\em Education Research International}, 2012(1):428639, 2012.

\bibitem{cleary2001}
Timothy~J. Cleary and Barry~J. Zimmerman.
\newblock Self-regulation differences during athletic competition: A microanalytic analysis.
\newblock {\em Journal of Applied Sport Psychology}, 13(2):185--206, 2001.

\bibitem{cloude2022system}
Elizabeth~Brooke Cloude, Roger Azevedo, Philip~H Winne, Gautam Biswas, and Eunice~E Jang.
\newblock System design for using multimodal trace data in modeling self-regulated learning.
\newblock In {\em Frontiers in Education}, volume~7, page 928632. Frontiers Media SA, 2022.

\bibitem{cohen1960coefficient}
Jacob Cohen.
\newblock A coefficient of agreement for nominal scales.
\newblock {\em Educational and Psychological Measurement}, 20(1):37--46, 1960.

\bibitem{darvishi2024impact}
Ali Darvishi, Hassan Khosravi, Shazia Sadiq, Dragan Ga{\v{s}}evi{\'c}, and George Siemens.
\newblock Impact of ai assistance on student agency.
\newblock {\em Computers \& Education}, 210:104967, 2024.

\bibitem{efklides2011interactions}
Anastasia Efklides.
\newblock Interactions of metacognition with motivation and affect in self-regulated learning: The masrl model.
\newblock {\em Educational psychologist}, 46(1):6--25, 2011.

\bibitem{efron1992bootstrap}
Bradley Efron.
\newblock Bootstrap methods: another look at the jackknife.
\newblock In {\em Breakthroughs in statistics: Methodology and distribution}, pages 569--593. Springer, 1992.

\bibitem{ekin2023prompt}
Sabit Ekin.
\newblock Prompt engineering for chatgpt: a quick guide to techniques, tips, and best practices.
\newblock {\em Authorea Preprints}, 2023.

\bibitem{fan2025beware}
Yizhou Fan, Luzhen Tang, Huixiao Le, Kejie Shen, Shufang Tan, Yueying Zhao, Yuan Shen, Xinyu Li, and Dragan Ga{\v{s}}evi{\'c}.
\newblock Beware of metacognitive laziness: Effects of generative artificial intelligence on learning motivation, processes, and performance.
\newblock {\em British Journal of Educational Technology}, 56(2):489--530, 2025.

\bibitem{flavell1979metacognition}
John~H. Flavell.
\newblock Metacognition and cognitive monitoring: A new area of cognitive–developmental inquiry.
\newblock {\em American Psychologist}, 34(10):906--911, 1979.

\bibitem{graham2000role}
Steve Graham and Karen R.~Harris.
\newblock The role of self-regulation and transcription skills in writing and writing development.
\newblock {\em Educational psychologist}, 35(1):3--12, 2000.

\bibitem{greene2011}
Jeffrey~A. Greene, Lara-Jeane Costa, and Kristin Dellinger.
\newblock Analysis of self-regulated learning processing using statistical models for count data.
\newblock {\em Metacognition and Learning}, 6(3):275--301, 2011.

\bibitem{greene2009macro}
Jeffrey~Alan Greene and Roger Azevedo.
\newblock A macro-level analysis of srl processes and their relations to the acquisition of a sophisticated mental model of a complex system.
\newblock {\em Contemporary educational psychology}, 34(1):18--29, 2009.

\bibitem{disco}
Christian~W G{\"u}nther and Anne Rozinat.
\newblock Disco: Discover your processes.
\newblock In {\em Demonstration Track of the 10th International Conference on Business Process Management, BPM Demos 2012}, pages 40--44. CEUR-WS. org, 2012.

\bibitem{fuzzyminer}
Christian~W G{\"u}nther and Wil~MP Van Der~Aalst.
\newblock Fuzzy mining--adaptive process simplification based on multi-perspective metrics.
\newblock In {\em International conference on business process management}, pages 328--343. Springer, 2007.

\bibitem{hacker1998metacognition}
Douglas~J Hacker, John Dunlosky, and Arthur~C Graesser.
\newblock {\em Metacognition in educational theory and practice}.
\newblock Routledge, 1998.

\bibitem{han-etal-2024-recipe4u-student}
Jieun Han, Haneul Yoo, Junho Myung, Minsun Kim, Tak~Yeon Lee, So-Yeon Ahn, and Alice Oh.
\newblock {RECIPE}4{U}: Student-{C}hat{GPT} interaction dataset in {EFL} writing education.
\newblock In {\em Proceedings of the 2024 Joint International Conference on Computational Linguistics, Language Resources and Evaluation (LREC-COLING 2024)}, pages 13666--13676, Torino, Italy, May 2024. ELRA and ICCL.

\bibitem{han2023exploring}
Jieun Han, Haneul Yoo, Junho Myung, Minsun Kim, Tak~Yeon Lee, So-Yeon Ahn, Alice Oh, and Acknowledgment~Negotiation Answer.
\newblock Exploring student-chatgpt dialogue in efl writing education.
\newblock In {\em 37th Conference on Neural Information Processing Systems. Neural Information Processing Systems Foundation}, 2023.

\bibitem{jarvela2024triggers}
Sanna J{\"a}rvel{\"a} and Allyson Hadwin.
\newblock Triggers for self-regulated learning: A conceptual framework for advancing multimodal research about srl.
\newblock {\em Learning and Individual Differences}, 115:102526, 2024.

\bibitem{jarvela2019capturing}
Sanna J{\"a}rvel{\"a}, Hanna J{\"a}rvenoja, and Jonna Malmberg.
\newblock Capturing the dynamic and cyclical nature of regulation: Methodological progress in understanding socially shared regulation in learning.
\newblock {\em International journal of computer-supported collaborative learning}, 14:425--441, 2019.

\bibitem{ji2024mobile}
Hyangeun Ji and Insook Han.
\newblock Mobile-based chatbot to scaffold foreign language learners' self-regulated learning.
\newblock {\em 교육방법연구}, 36(1):67--88, 2024.

\bibitem{jarvela2021}
Sanna Järvelä, Päivi H{\"a}m{\"a}l{\"a}inen, Allyson~F. Hadwin, and Sanna Malmberg.
\newblock What multimodal data can tell us about the students’ regulation of learning?
\newblock {\em Learning and Instruction}, 72:101219, 2021.

\bibitem{kim2022perceived}
Jihyun Kim, Kelly Merrill~Jr, Kun Xu, and Stephanie Kelly.
\newblock Perceived credibility of an ai instructor in online education: The role of social presence and voice features.
\newblock {\em Computers in Human Behavior}, 136:107383, 2022.

\bibitem{kuhail2023interacting}
Mohammad~Amin Kuhail, Nazik Alturki, Salwa Alramlawi, and Kholood Alhejori.
\newblock Interacting with educational chatbots: A systematic review.
\newblock {\em Education and Information Technologies}, 28(1):973--1018, 2023.

\bibitem{lee2024empowering}
Hsin-Yu Lee, Pei-Hua Chen, Wei-Sheng Wang, Yueh-Min Huang, and Ting-Ting Wu.
\newblock Empowering chatgpt with guidance mechanism in blended learning: Effect of self-regulated learning, higher-order thinking skills, and knowledge construction.
\newblock {\em International Journal of Educational Technology in Higher Education}, 21(1):16, 2024.

\bibitem{leung2022mapping}
Javier Leung.
\newblock Mapping self-regulated learning events and actions in online teacher professional development with process mining techniques.
\newblock {\em Quarterly Review of Distance Education}, 23(4):31--62, 2022.

\bibitem{li2025facilitating}
Tongguang Li.
\newblock {\em Facilitating Analytics-based Adaptive Scaffolding for Self-regulated Learning}.
\newblock PhD thesis, Monash University, 2025.

\bibitem{li2023analytics}
Tongguang Li, Yizhou Fan, Yuanru Tan, Yeyu Wang, Shaveen Singh, Xinyu Li, Mladen Rakovi{\'c}, Joep Van Der~Graaf, Lyn Lim, Binrui Yang, et~al.
\newblock Analytics of self-regulated learning scaffolding: effects on learning processes.
\newblock {\em Frontiers in Psychology}, 14:1206696, 2023.

\bibitem{lin2023chat}
Michael Pin-Chuan Lin and Daniel Chang.
\newblock Chat-acts: A pedagogical framework for personalized chatbot to enhance active learning and self-regulated learning.
\newblock {\em Computers and Education: Artificial Intelligence}, 5:100167, 2023.

\bibitem{long2014penetrating}
Phillip Long and George Siemens.
\newblock Penetrating the fog: analytics in learning and education.
\newblock {\em Italian Journal of Educational Technology}, 22(3):132--137, 2014.

\bibitem{luckin2016intelligence}
Rose Luckin, Wayne Holmes, Mark Griffiths, and Laurene~B. Forcier.
\newblock {\em Intelligence Unleashed: An Argument for AI in Education}.
\newblock Pearson Education, London, 2016.
\newblock Accessed May 2025.

\bibitem{luo2024effectiveness}
Ren-Zhi Luo and Yue-Liang Zhou.
\newblock The effectiveness of self-regulated learning strategies in higher education blended learning: A five years systematic review.
\newblock {\em Journal of Computer Assisted Learning}, 40(6):3005--3029, 2024.

\bibitem{lyu2024mamba}
Yilin Lyu.
\newblock A mamba-based approximate conformance checking method.
\newblock In {\em Fourth International Conference on Advanced Algorithms and Neural Networks (AANN 2024)}, volume 13416, pages 885--891. SPIE, 2024.

\bibitem{lyu2024discussion}
Yilin Lyu and Bo~Yin.
\newblock A discussion of migration of common neural network regularization methods on snns.
\newblock In {\em Ninth International Symposium on Advances in Electrical, Electronics, and Computer Engineering (ISAEECE 2024)}, volume 13291, pages 1355--1361. SPIE, 2024.

\bibitem{maghfiroh2017problem}
Lailiyah Maghfiroh, Wachju Subchan, and M~Iqbat.
\newblock Problem based learning through moodle for increasing self-regulated learning students (goal setting and planning).
\newblock {\em The International Journal of Social Sciences and Humanities Investigation}, 4(8):3880--3887, 2017.

\bibitem{maldonado2022miranda}
Jorge Maldonado-Mahauad, Mar P{\'e}rez-Sanagust{\'\i}n, Juan Carvallo-Vega, Edwin Narvaez, and Mauricio Calle.
\newblock Miranda: A chatbot for supporting self-regulated learning.
\newblock In {\em European conference on technology enhanced learning}, pages 455--462. Springer, 2022.

\bibitem{molenaar2014advances}
Inge Molenaar.
\newblock Advances in temporal analysis in learning and instruction.
\newblock {\em Frontline Learning Research}, 2(4):15--24, 2014.

\bibitem{molenaar2022a}
Inge Molenaar.
\newblock Co-regulation with ai: A developmental perspective on hybrid human-ai regulation in education.
\newblock In {\em Proceedings of the 15th International Conference of the Learning Sciences (ICLS)}, pages 1033--1040, 2022.

\bibitem{molenaar2023measuring}
Inge Molenaar, Susanne de~Mooij, Roger Azevedo, Maria Bannert, Sanna J{\"a}rvel{\"a}, and Dragan Ga{\v{s}}evi{\'c}.
\newblock Measuring self-regulated learning and the role of ai: Five years of research using multimodal multichannel data.
\newblock {\em Computers in Human Behavior}, 139:107540, 2023.

\bibitem{mooij2014self}
Ton Mooij, Karl Steffens, and Maureen~Snow Andrade.
\newblock Self-regulated and technology-enhanced learning: A european perspective, 2014.

\bibitem{munshi2022analyzing}
Anabil Munshi, Gautam Biswas, Ryan Baker, Jaclyn Ocumpaugh, Stephen Hutt, and Luc Paquette.
\newblock Analyzing adaptive scaffolds that help students develop self-regulated learning behaviors.
\newblock {\em Journal of Computer Assisted Learning}, 38(6):1420--1435, 2022.

\bibitem{osakwe2024measurement}
Ikenna Osakwe, Guanliang Chen, Yizhou Fan, Mladen Rakovic, Shaveen Singh, Inge Molenaar, and Dragan Ga{\v{s}}evi{\'c}.
\newblock Measurement of self-regulated learning: Strategies for mapping trace data to learning processes and downstream analysis implications.
\newblock In {\em Proceedings of the 14th Learning Analytics and Knowledge Conference}, pages 563--575, 2024.

\bibitem{panadero2017}
Ernesto Panadero.
\newblock A review of self-regulated learning: Six models and four directions for research.
\newblock {\em Frontiers in Psychology}, 8:422, 2017.

\bibitem{panadero2017review}
Ernesto Panadero.
\newblock A review of self-regulated learning: Six models and four directions for research.
\newblock {\em Frontiers in Psychology}, 8:422, 2017.

\bibitem{paul2019thinker}
Richard Paul and Linda Elder.
\newblock {\em The thinker's guide to Socratic questioning}.
\newblock Rowman \& Littlefield, 2019.

\bibitem{pintrich2000role}
Paul~R Pintrich.
\newblock The role of goal orientation in self-regulated learning.
\newblock In {\em Handbook of self-regulation}, pages 451--502. Elsevier, 2000.

\bibitem{pintrich2004}
Paul~R. Pintrich.
\newblock A conceptual framework for assessing motivation and self-regulated learning in college students.
\newblock {\em Educational Psychology Review}, 16(4):385--407, 2004.

\bibitem{pintrich1991manual}
Paul~R Pintrich et~al.
\newblock A manual for the use of the motivated strategies for learning questionnaire (mslq).
\newblock 1991.

\bibitem{puustinen2001models}
Minna Puustinen and Lea Pulkkinen.
\newblock Models of self-regulated learning: A review.
\newblock {\em Scandinavian journal of educational research}, 45(3):269--286, 2001.

\bibitem{Puustinen2001}
Minna Puustinen and Lea Pulkkinen.
\newblock Models of self-regulated learning: A review.
\newblock {\em Scandinavian Journal of Educational Research}, 45(3):269--286, 2001.

\bibitem{rienties2016implementing}
Bart Rienties, Simon Cross, and Zdenek Zdrahal.
\newblock Implementing a learning analytics intervention and evaluation framework: What works?
\newblock In {\em Big data and learning analytics in higher education: Current theory and practice}, pages 147--166. Springer, 2016.

\bibitem{risko2016}
Evan~F. Risko and Sam~J. Gilbert.
\newblock Cognitive offloading.
\newblock {\em Trends in Cognitive Sciences}, 20(9):676--688, 2016.

\bibitem{roll2011helpseeking}
Ido Roll, Vincent Aleven, Bruce~M. McLaren, and Kenneth~R. Koedinger.
\newblock Improving students’ help-seeking skills using metacognitive feedback in an intelligent tutoring system.
\newblock {\em Learning and Instruction}, 21(2):267--280, 2011.

\bibitem{roll2011metacognitive}
Ido Roll, Vincent Aleven, Bruce~M McLaren, and Kenneth~R Koedinger.
\newblock Metacognitive support promotes transfer: Nudge, don't shove.
\newblock {\em International Journal of Artificial Intelligence in Education}, 21(1-2):1--28, 2011.

\bibitem{romero2010educational}
Cristobal Romero, Sebastian Ventura, Mykola Pechenizkiy, and Ryan~SJ Baker.
\newblock Educational data mining: A review of the state of the art.
\newblock {\em IEEE Transactions on Systems, Man, and Cybernetics, Part C (Applications and Reviews)}, 40(6):601--618, 2010.

\bibitem{rovers2019granularity}
Sanne~FE Rovers, Geraldine Clarebout, Hans~HCM Savelberg, Anique~BH De~Bruin, and Jeroen~JG van Merri{\"e}nboer.
\newblock Granularity matters: comparing different ways of measuring self-regulated learning.
\newblock {\em Metacognition and Learning}, 14:1--19, 2019.

\bibitem{saint2020combining}
John Saint, Dragan Ga{\v{s}}evi{\'c}, Wannisa Matcha, Nora'Ayu~Ahmad Uzir, and Abelardo Pardo.
\newblock Combining analytic methods to unlock sequential and temporal patterns of self-regulated learning.
\newblock In {\em Proceedings of the tenth international conference on learning analytics \& knowledge}, pages 402--411, 2020.

\bibitem{saint2020tracesrl}
John Saint, Alexander Whitelock-Wainwright, Dragan Ga\v{s}evi\'{c}, and Abelardo Pardo.
\newblock Trace-{SRL}: A framework for analysis of micro-level processes of self-regulated learning from trace data.
\newblock {\em IEEE Transactions on Learning Technologies}, 13(4):861--877, 2020.

\bibitem{SchunkZimmerman2012}
Dale~H. Schunk and Barry~J. Zimmerman.
\newblock Self-regulated learning: Theories, measures, and outcomes.
\newblock In Barry~J. Zimmerman and Dale~H. Schunk, editors, {\em Handbook of Self-Regulation of Learning and Performance}, pages 1--24. Routledge, 2012.

\bibitem{she2023learning}
Chunmei She, Qiao Liang, Wenjun Jiang, and Qiang Xing.
\newblock Learning adaptability facilitates self-regulated learning at school: the chain mediating roles of academic motivation and self-management.
\newblock {\em Frontiers in psychology}, 14:1162072, 2023.

\bibitem{siadaty2016}
Mehrdad Siadaty, Dragan Ga{\v{s}}evi{\'c}, Jelena Jovanovi{\'c}, Abelardo Pardo, and Shane Dawson.
\newblock Analytics of learners’ self-regulated learning tactics in a social, collaborative learning environment.
\newblock {\em Computers in Human Behavior}, 55:550--567, 2016.

\bibitem{siadaty2016trace}
Melody Siadaty, Dragan Gasevic, and Marek Hatala.
\newblock Trace-based micro-analytic measurement of self-regulated learning processes.
\newblock {\em Journal of Learning Analytics}, 3(1):183--214, 2016.

\bibitem{sobocinski2017exploring}
M{\'a}rta Sobocinski, Jonna Malmberg, and Sanna J{\"a}rvel{\"a}.
\newblock Exploring temporal sequences of regulatory phases and associated interactions in low-and high-challenge collaborative learning sessions.
\newblock {\em Metacognition and Learning}, 12:275--294, 2017.

\bibitem{sonnenberg2015discovering}
Christoph Sonnenberg and Maria Bannert.
\newblock Discovering the effects of metacognitive prompts on the sequential structure of srl-processes using process mining techniques.
\newblock {\em Journal of Learning Analytics}, 2(1):72--100, 2015.

\bibitem{sulisworo2020students}
Dwi Sulisworo, Meita Fitrianawati, Ika Maryani, Saleh Hidayat, Erie Agusta, and Wulandari Saputri.
\newblock Students' self-regulated learning (srl) profile dataset measured during covid-19 mitigation in yogyakarta, indonesia.
\newblock {\em Data in Brief}, 33:106422, 2020.

\bibitem{tempelaar2024understanding}
Dirk Tempelaar, Anik{\'o} B{\'a}tori, and Bas Giesbers.
\newblock Understanding self-regulation strategies in problem-based learning through dispositional learning analytics.
\newblock In {\em Frontiers in Education}, volume~9, page 1382771. Frontiers Media SA, 2024.

\bibitem{urban2024chatgpt}
Marek Urban, Filip D{\v{e}}cht{\v{e}}renko, Ji{\v{r}}{\'\i} Lukavsk{\`y}, Veronika Hrabalov{\'a}, Filip Svacha, Cyril Brom, and Kamila Urban.
\newblock Chatgpt improves creative problem-solving performance in university students: An experimental study.
\newblock {\em Computers \& Education}, 215:105031, 2024.

\bibitem{van2012process1}
Wil Van Der~Aalst.
\newblock Process mining: Overview and opportunities.
\newblock {\em ACM Transactions on Management Information Systems (TMIS)}, 3(2):1--17, 2012.

\bibitem{van2016data}
Wil Van Der~Aalst and Wil van~der Aalst.
\newblock {\em Data science in action}.
\newblock Springer, 2016.

\bibitem{van2004workflow}
Wil Van~der Aalst, Ton Weijters, and Laura Maruster.
\newblock Workflow mining: Discovering process models from event logs.
\newblock {\em IEEE transactions on knowledge and data engineering}, 16(9):1128--1142, 2004.

\bibitem{vanderAalst2016}
Wil M.~P. van~der Aalst.
\newblock {\em Process Mining: Data Science in Action}.
\newblock Springer, 2 edition, 2016.

\bibitem{vandevelde2015using}
Sabrina Vandevelde, Hilde Van~Keer, Gonny Schellings, and Bernadette Van Hout-Wolters.
\newblock Using think-aloud protocol analysis to gain in-depth insights into upper primary school children's self-regulated learning.
\newblock {\em Learning and Individual Differences}, 43:11--30, 2015.

\bibitem{walter2024embracing}
Yoshija Walter.
\newblock Embracing the future of artificial intelligence in the classroom: the relevance of ai literacy, prompt engineering, and critical thinking in modern education.
\newblock {\em International Journal of Educational Technology in Higher Education}, 21(1):15, 2024.

\bibitem{winne2010framework}
Philip~H. Winne.
\newblock A metacognitive view of individual differences in self-regulated learning.
\newblock {\em Educational Psychologist}, 45(4):267--276, 2010.

\bibitem{winne2011cognitive}
Philip~H Winne.
\newblock A cognitive and metacognitive analysis of self-regulated learning: Faculty of education, simon fraser university, burnaby, canada.
\newblock In {\em Handbook of self-regulation of learning and performance}, pages 29--46. Routledge, 2011.

\bibitem{winne2017learning}
Philip~H Winne.
\newblock Learning analytics for self-regulated learning.
\newblock {\em Handbook of learning analytics}, 754:241--249, 2017.

\bibitem{winters2008self}
Fielding~I Winters, Jeffrey~A Greene, and Claudine~M Costich.
\newblock Self-regulation of learning within computer-based learning environments: A critical analysis.
\newblock {\em Educational psychology review}, 20:429--444, 2008.

\bibitem{xu2025enhancing}
Xiaoqing Xu, Lifang Qiao, Nuo Cheng, Hongxia Liu, and Wei Zhao.
\newblock Enhancing self-regulated learning and learning experience in generative ai environments: The critical role of metacognitive support.
\newblock {\em British Journal of Educational Technology}, 2025.

\bibitem{zimmerman1989social}
Barry~J Zimmerman.
\newblock A social cognitive view of self-regulated academic learning.
\newblock {\em Journal of educational psychology}, 81(3):329, 1989.

\bibitem{zimmerman2000attaining}
Barry~J Zimmerman.
\newblock Attaining self-regulation: A social cognitive perspective.
\newblock In {\em Handbook of self-regulation}, pages 13--39. Elsevier, 2000.

\bibitem{Zimmerman2008}
Barry~J. Zimmerman.
\newblock {\em Investigating self-regulation and motivation: Historical background, methodological developments, and future prospects}, volume~45.
\newblock 2008.

\end{thebibliography}
\end{CJK}

\newpage
\appendix
\renewcommand{\thetable}{A.\arabic{table}}
\section*{Appendix A: Prompt–Action Dictionary}
\label{appendix:dictionary}

\begin{longtable}{|p{5cm}|p{10cm}|}
\caption{Prompt--Action Dictionary}\\
\hline
\textbf{Action} & \textbf{Example Prompts} \\
\hline
\endfirsthead

\multicolumn{2}{l}%
{{\bfseries \tablename\ \thetable{} -- continued from previous page}} \\
\hline
\textbf{Action} & \textbf{Example Prompts} \\
\hline
\endhead

\hline \multicolumn{2}{|r|}{{Continued on next page}} \\ \hline
\endfoot
\hline
\endlastfoot

\multirow{10}{*}{}{Define Problem} & Today I need to solve ... \\\cline{2-2}
          & I don’t get the requirement of this task … \\\cline{2-2}
          & I’m stuck at the beginning … \\\cline{2-2}
          & I don’t know how to … \\\cline{2-2}
          & The assignment topic is confusing … \\\cline{2-2}
          & For today's lecture, I’m not sure about … \\\cline{2-2}
          & I’m confused about … \\\cline{2-2}
          & The issue I have is … \\\cline{2-2}
          & I’m trying to understand … \\\cline{2-2}
          & … doesn’t make sense to me \\ \hline

\multirow{10}{*}{}{Set Goals}
   & I want to understand/write ... \\\cline{2-2}
          & My goal is to improve ... \\\cline{2-2}
          & Today/This time, I plan to work on ... \\\cline{2-2}
          & I want to master ... \\\cline{2-2}
          & I will focus on ... \\\cline{2-2}
          & I aim to ... \\\cline{2-2}
          & I’m setting a goal to ... \\\cline{2-2}
          & I need to achieve ... \\ \hline

\multirow{10}{*}{}{Check Understanding - Basic}&I see. That means ... \\\cline{2-2}
          & So are you saying ...? \\\cline{2-2}
          & I think you mean ... \\\cline{2-2}
          & If I understand correctly, ... \\\cline{2-2}
          & Just to confirm, ...? \\\cline{2-2}
          & Does that mean ...? \\\cline{2-2}
          & Do you think (this understanding)... is right/wrong? \\\cline{2-2}
          & So, in other words ...? \\\cline{2-2}
          & Am I right in ...? \\\cline{2-2}
          & You’re saying ...? \\\cline{2-2}
          & This means that ...? \\\hline

\multirow{14}{*}{}{Check Understanding - Deep}
   & If … is …, then can it still be considered …? \\\cline{2-2}
          & So this is like …, right? \\\cline{2-2}
          & Would this also apply if I were doing … instead of …? \\\cline{2-2}
          & Is there a case where this wouldn’t work? \\\cline{2-2}
          & Is this like how … works in …? \\\cline{2-2}
          & Can I think of this as similar to …? \\\cline{2-2}
          & So it is different from …? \\\cline{2-2}
          & What happens if we change …? \\\cline{2-2}
          & Could this be used in another context? \\\cline{2-2}
          & How does this compare with (another field)…? \\\cline{2-2}
          & Would a counterexample be …? \\\cline{2-2}
          & Would this help when I am doing (another field)...? \\\cline{2-2}
          & Does this rule out …? \\\cline{2-2}
          & Could you apply this to another situation? \\ \hline

\multirow{10}{*}{}{Seek Clarification - Basic}
   & Can you clarify the concept you just mentioned? \\\cline{2-2}
          & Can you make it easier to understand? \\\cline{2-2}
          & What do you mean by ...? \\\cline{2-2}
          & Make it clear ... \\\cline{2-2}
          & Make it detailed ... \\\cline{2-2}
          & Could you explain ... again? \\\cline{2-2}
          & Give me some simple explanations for ... \\\cline{2-2}
          & Can you simplify this concept...? \\\cline{2-2}
          & How to explain the concept of ...? \\\cline{2-2}
          & Can you rephrase ... in a simple way? \\\cline{2-2}
          & What is the definition of ...? \\\cline{2-2}
          & How do you define ...? \\\cline{2-2}
          & What does ... mean? \\ \hline

\multirow{11}{*}{}{Seek Clarification - Deep}
    & Could you walk me through an example about this concept? \\ \cline{2-2}
    & Can you give an example for this concept? \\ \cline{2-2}
    & Why is this concept better than the others? \\ \cline{2-2}
    & What is the difference between concept A and B? \\ \cline{2-2}
    & Can you show how this concept works in practice? \\ \cline{2-2}
    & What would be a counterexample of this concept? \\ \cline{2-2}
    & Can you compare this concept with ...? \\ \cline{2-2}
    & Can you provide a case about this concept where ...? \\ \cline{2-2}
    & What causes this concept ...? \\ \cline{2-2}
    & Explain this concept in a different tone. \\ \cline{2-2}
    & What are the implications of this concept ...? \\ \cline{2-2}
    & Can you contrast this concept with (another concept)...? \\ \hline

\multirow{10}{*}{}{Request Help - Instrumental}
    & How can I make this paragraph longer? \\ \cline{2-2}
    & How can I re-structure the following sentence? \\ \cline{2-2}
    & Can you suggest any improvements on ...? \\ \cline{2-2}
    & Give me a outline of ... \\ \cline{2-2}
    & How should I organize my ideas? \\ \cline{2-2}
    & What would be a better way to write ...? \\ \cline{2-2}
    & Can you recommend a structure for ...? \\ \cline{2-2}
    & What's the best way to introduce ...? \\ \cline{2-2}
    & How do I improve the flow of ...? \\ \cline{2-2}
    & Can you help me polish ...? \\ \cline{2-2}
    & How do I connect these ideas? \\ \hline

\multirow{10}{*}{}{Request Help - Executive}
    & Please generate a new introduction for this essay. \\ \cline{2-2}
    & Write me a summary paragraph/draft. \\ \cline{2-2}
    & Can you write the conclusion for me? \\ \cline{2-2}
    & Can you rewrite this in a better way? \\ \cline{2-2}
    & Just give me a version of ... \\ \cline{2-2}
    & Can you do this for me? \\ \cline{2-2}
    & Produce a revised version of ... \\ \cline{2-2}
    & Please complete this sentence: \\ \cline{2-2}
    & Can you handle the rest part for me? \\ \cline{2-2}
    & Draft a paragraph about ... \\ \cline{2-2}
    & Give me the answer about ... \\ \cline{2-2}
    & Can you answer this question for me? \\ \hline

\multirow{10}{*}{}{Request Help - Check}
    & Can you help me revise/check ...? \\ \cline{2-2}
    & How would you rate ...? \\ \cline{2-2}
    & Is this written correctly? \\ \cline{2-2}
    & Can you review this ...? \\ \cline{2-2}
    & Does this make sense? \\ \cline{2-2}
    & What would you change in ...? \\ \cline{2-2}
    & Evaluate this for me: \\ \cline{2-2}
    & Does this sentence need fixing? \\ \cline{2-2}
    & Can you mark/rate my writing? \\ \cline{2-2}
    & How can I improve this part? \\ \hline

\multirow{13}{*}{}{Access Information}
    & What is ...(non-concept question)? \\ \cline{2-2}
    & Can you give some references about ...? \\ \cline{2-2}
    & Can you translate it into English? \\ \cline{2-2}
    & Can you count the words? \\ \cline{2-2}
    & Give me the BibTex format citation. \\ \cline{2-2}
    & Translate this citation into APA format. \\ \cline{2-2}
    & When was ... created? \\ \cline{2-2}
    & What's the meaning of ...? \\ \cline{2-2}
    & Tell me the author's background. \\ \cline{2-2}
    & How many words are there? \\ \cline{2-2}
    & What are the types of ...? \\ \cline{2-2}
    & What's the synonym of ...? \\ \cline{2-2}
    & What is the correct spelling of ...? \\ \cline{2-2}
    & What is your version ...? \\ \hline

\multirow{10}{*}{}{Refine Prompt}
    & Instead of just ..., can you ...? \\ \cline{2-2}
    & Let me rephrase that ... in a simple way. \\ \cline{2-2}
    & I meant to ask ... \\ \cline{2-2}
    & I'll try asking more clear ... \\ \cline{2-2}
    & Let's revise/change the question to ... \\ \cline{2-2}
    & Actually, can you ... instead? \\ \cline{2-2}
    & Sorry, what I meant was ... \\ \cline{2-2}
    & Let's try this version: \\ \cline{2-2}
    & Let's reword it like ... \\ \cline{2-2}
    & How about we try asking ... \\ \hline

\multirow{10}{*}{}{Correct Answer}
    & You made a mistake, I think the correct answer is ... \\ \cline{2-2}
    & Actually, that's incorrect because ... \\ \cline{2-2}
    & I believe the right explanation should be ... \\ \cline{2-2}
    & Let me correct that ... \\ \cline{2-2}
    & That's not accurate, the fact is ... \\ \cline{2-2}
    & You got it wrong, it should be ... \\ \cline{2-2}
    & Correction: ... \\ \cline{2-2}
    & That answer is outdated, the latest info is ... \\ \cline{2-2}
    & This response is misleading, give me a better answer. \\ \cline{2-2}
    & Your answer is not good enough, the truth/fact is ... \\ \hline

\multirow{10}{*}{}{Self-Evaluate}
    & I feel confident about ... but still struggle with ... \\ \cline{2-2}
    & What is the best way for me to ...? \\ \cline{2-2}
    & I did well on ... but need more help with ... \\ \cline{2-2}
    & I think I've improved in ... \\ \cline{2-2}
    & I'm satisfied with ... but not with ... \\ \cline{2-2}
    & I realized I still don't know how to ... \\ \cline{2-2}
    & I should spend more time on ... \\ \cline{2-2}
    & My strategy worked well for ... \\ \hline

\multirow{10}{*}{}{Summary Learning}
    & Today I learned... \\ \cline{2-2}
    & What I got from this session is ... \\ \cline{2-2}
    & I now understand ... \\ \cline{2-2}
    & This helped me learn ... \\ \cline{2-2}
    & From this, I've figured out ... \\ \cline{2-2}
    & I discovered that ... \\ \cline{2-2}
    & The key takeaway today is ... \\ \cline{2-2}
    & This session clarified/answered ... \\ \cline{2-2}
    & I've gained a better grasp of ... \\ \hline

\multirow{10}{*}{}{Plan Next Step}
    & I will try to ... in the next stage. \\ \cline{2-2}
    & Next, I plan to ... \\ \cline{2-2}
    & My next step is ... \\ \cline{2-2}
    & I'll review ... tomorrow. \\ \cline{2-2}
    & I'll go back and work on ... \\ \cline{2-2}
    & In my next session, I'll ... \\ \cline{2-2}
    & Next time I should ... \\ \cline{2-2}
    & I'll practice ... more. \\ \cline{2-2}
    & I need to revisit ... later. \\ \cline{2-2}
    & I'll revise my ... next. \\ \cline{2-2}
    & I should focus on ... next time \\ \hline

\multirow{8}{*}{}{Acknowledgment}
    & Thanks! \\ \cline{2-2}
    & Morning. \\ \cline{2-2}
    & Okay, great. \\ \cline{2-2}
    & Cool! \\ \cline{2-2}
    & I appreciate it. \\ \cline{2-2}
    & Sounds good. \\ \cline{2-2}
    & Bye! \\ \cline{2-2}
    & Take care. \\ \hline

\end{longtable}

\appendix
\renewcommand{\thetable}{B.\arabic{table}} 
\setcounter{table}{0}

\section*{Appendix B: Transition Frequency Matrix}
\label{appendix:transition frequency matrix}

This appendix includes the full transition frequency matrix supporting the process models in Section~\ref{subsubsec:process analysis}.

\begin{table}[htbp]
\centering
\caption{Transition frequency matrix of annotated macrolevel phases in the RECIPE4U dataset ($N=894$), ordered by SRL phase sequence, showing phase frequencies and their ordered transitions. Read the row first.}
\label{tab:macro_transition_matrix}
\begin{tabular}{l|rrrr}
\toprule
 & Forethought & Monitoring & Control & Reflection \\
N (Total)     & 48 & 97 & 733 & 16 \\
\midrule
Forethought & 8  & 5  & 25 & 2 \\
Monitoring  & 8  & 31 & 53 & 0 \\
Control     & 30 & 57 & 610 & 8 \\
Reflection  & 1  & 3  & 9  & 2 \\
\bottomrule
\end{tabular}
\end{table}

\begin{table}[htbp]
\centering
\caption{Transition frequency matrix of annotated SRL actions in the RECIPE4U dataset ($N=894$), showing action frequencies and their ordered transitions. Read the row first.}
\label{tab:activity_recipe4u}
\begin{tabular}{l|rrrrrrrrrrrrrrr}
\toprule
 & F.DP & F.SG & M.CU(B) & M.CU(D) & C.SC(B) & C.SC(D) & C.RH(I) & C.RH(E) & C.RH(C) & C.AI & C.RP & C.CA & R.SE & R.SL & R.PN \\
N & 27 & 21 & 79 & 16 & 91 & 43 & 143 & 145 & 184 & 97 & 17 & 16 & 4 & 3 & 8 \\
\midrule
F.DP & 6 & 1 & 4 & 0 & 2 & 2 & 2 & 3 & 5 & 1 & 0 & 0 & 0 & 0 & 0 \\
F.SG & 1 & 0 & 4 & 0 & 0 & 0 & 3 & 4 & 5 & 3 & 0 & 0 & 0 & 1 & 0 \\
M.CU(B) & 3 & 2 & 27 & 2 & 7 & 4 & 3 & 8 & 11 & 5 & 1 & 3 & 2 & 0 & 0 \\
M.CU(D) & 0 & 0 & 1 & 2 & 3 & 2 & 1 & 2 & 1 & 4 & 0 & 0 & 0 & 0 & 0 \\
C.SC(B) & 2 & 1 & 3 & 0 & 28 & 6 & 14 & 9 & 11 & 10 & 0 & 2 & 0 & 0 & 0 \\
C.SC(D) & 1 & 0 & 4 & 1 & 8 & 7 & 8 & 4 & 5 & 4 & 0 & 0 & 0 & 0 & 0 \\
C.RH(I) & 1 & 5 & 9 & 4 & 8 & 8 & 36 & 26 & 17 & 16 & 2 & 1 & 1 & 0 & 4 \\
C.RH(E) & 5 & 1 & 9 & 0 & 7 & 7 & 25 & 42 & 20 & 9 & 7 & 2 & 0 & 0 & 1 \\
C.RH(C) & 3 & 2 & 7 & 3 & 12 & 2 & 30 & 24 & 71 & 14 & 3 & 3 & 1 & 1 & 2 \\
C.AI & 0 & 2 & 3 & 2 & 10 & 1 & 17 & 14 & 13 & 23 & 1 & 4 & 0 & 0 & 0 \\
C.RP & 0 & 1 & 2 & 0 & 3 & 0 & 1 & 1 & 5 & 2 & 0 & 0 & 0 & 0 & 0 \\
C.CA & 0 & 1 & 2 & 2 & 3 & 0 & 0 & 3 & 1 & 2 & 1 & 1 & 0 & 0 & 0 \\
R.SE & 0 & 0 & 0 & 0 & 0 & 1 & 1 & 1 & 1 & 0 & 0 & 0 & 0 & 0 & 0 \\
R.SL & 1 & 1 & 0 & 0 & 0 & 0 & 0 & 0 & 1 & 0 & 0 & 0 & 0 & 0 & 0 \\
R.PN & 0 & 0 & 0 & 0 & 0 & 0 & 2 & 0 & 1 & 0 & 0 & 0 & 0 & 0 & 1 \\
\bottomrule
\end{tabular}
\end{table}

\end{document}